\documentclass{aa}  

\usepackage{graphicx}  
\usepackage{txfonts}
\usepackage{hyperref}
\usepackage{orcidlink}
\usepackage[switch]{lineno}
\newcommand{\teff}{\hbox{$T_{\rm eff}$}}
\newcommand{\logg}{\hbox{log\,$\it g$}}
\newcommand{\feh}{\hbox{$\rm [Fe/H]$}}
\newcommand{\absm}{\hbox{$M_{\rm G}$}}
\newcommand{\bprp}{\hbox{$(\rm{{BP} - {RP}})_{0}$}}
\usepackage{color}
\definecolor{black}{rgb}{0,0,0}
\definecolor{blue}{rgb}{0,0,1}

\begin{document} 

\title{Distance and stellar parameter estimations of solar-like stars from the LAMOST spectroscopic survey}
\titlerunning{Distance and stellar parameter estimations of solar-like stars}

	\author{Yue-Yue Shen\inst{1,2}\orcidlink{0000-0001-8636-2990}
			\and
			A-Li Luo\inst{1,2}\fnmsep\thanks{A-Li Luo  ~~  email: lal@nao.cas.cn}
			\orcidlink{0000-0001-7865-2648}
			}
	\authorrunning{Yue-Yue Shen and A-Li Luo}
	
	\institute{CAS Key Laboratory of Optical Astronomy, 
				National Astronomical Observatories, Beijing 100101, China\\
				\and
				School of Astronomy and Space Science, University of Chinese Academy of Sciences, Beijing 100049, China\\
	}

	\date{Received mm:dd, yyyy; accepted mm:dd, yyyy}
 
	\abstract
	 {%
	 The Gaia mission has opened up a new era for the precise astrometry of stars, thus revolutionizing our understanding of the Milky Way. However, beyond a few kiloparseconds from the Sun, parallax measurements become less reliable, and even within 2 kpc, there still exist stars with large uncertainties.
	 }%
	 {%
	 Our aim was to determine the distance and stellar parameters of 521,424 solar-like stars from LAMOST DR9; these stars  lacked precise distance measurements (uncertainties higher than 20\% or even without any distance estimations) when checked with Gaia.
	 }%
	 {%
	 We proposed a convolutional neural network (CNN) model to predict the absolute magnitudes, colors, and stellar parameters (\teff, \logg, and \feh) directly from low-resolution spectra. For spectra with signal-to-noise ratios at $g$ band ($\rm{S/N_g}$) greater than 10, the model achieves a precision of 85 K for \teff, 0.07 dex for \logg, 0.06 dex for \feh, 0.25 mag for \absm, and 0.03 mag for \bprp. The estimated distances have a median fractional error of 4\% with a standard deviation of 8\%.
	 }%
	 {%
	 We applied the trained CNN model to 521,424 solar-like stars to derive the distance and stellar parameters. Compared with other distance estimation studies and spectroscopic surveys, the results show good consistency. Additionally, we investigated the metallicity gradients of the Milky Way from a subsample, and find a radial gradient ranging from $-0.05 < \Delta \feh / \Delta R < 0.0\ {\rm{dex\ kpc^{-1}}}$ and a vertical gradient ranging from $-0.26 < \Delta \feh / \Delta Z < -0.07\ \rm{dex\ kpc^{-1}}$.
	 }%
	 {%
	 We conclude that our method is effective in estimating distances and stellar parameters for solar-like stars with limited astrometric data. Our measurements are reliable for Galactic structure studies and hopefully will be useful for exoplanet researches.
	 }%


   \maketitle
%
\section{Introduction}
\label{sec:info}
As we know, planet-hunting missions such as the Convection, Rotation and planetary Transits 
(CoRoT; \citealp{baglin_corot_2006, borde_exoplanet_2003}), Kepler \citep{borucki_kepler_2010}, 
and the Transiting Exoplanet Survey Satellite (TESS; \citealp{Ricker_transiting_2016}) 
have revealed that solar-like stars commonly harbor exoplanets. 
Approximately one-half of these stars are observed to possess rocky planets located within their habitable zones.  
The distance of solar-like stars significantly influences exoplanet searches, affecting their detectability. 
Bright nearby solar-like stars represent the most promising targets for exoplanet detection, 
necessitating follow-up observations using diverse methods.

With the accurate Gaia parallax measurements \citep{bailer-jones_estimating_2021}, their accurate luminosities can thus be derived. However, for stars at larger distances, 
accurate parallaxes are more difficult to measure; for example, at 5 kpc the median error in Gaia data is 15\%, and at 8 kpc it grows to 34\% \citep{gaiaDR3summary}. Previous work has used the combination of astrometric, photometric, and spectroscopic data along with different methods to determine distances (see, e.g., \citealt{hawley_characterization_2002}, \citealt{burnett_stellar_2010}, \citealt{santiago_spectro-photometric_2016}, \citealt{queiroz_starhorse_2018}, \citealt{das_made_2019}, \citealt{leung_deep_2019}, \citealt{stone-martinez_spectroscopic_2024}). 
\citet{santiago_spectro-photometric_2016} developed a Bayesian inference code to determine spectro-photometric distances, followed by \citet{queiroz_starhorse_2018}, who presented \texttt{StarHorse} with extinctions and other parameters added, and improved the code with extinction treatment in \citet{queiroz_bulge_2020}. Instead of pure Bayesian inference, \citet{das_made_2019} trained \texttt{MADE}, a Bayesian artificial neural network (ANN) that learns from and replaces stellar isochrones to estimate distance. 

Beyond the reach of Gaia's parallax, we can use the relation between spectra and luminosity (or absolute magnitude) to determine distances. Meanwhile, thanks to the large amount of stars with precise parallaxes that Gaia has already provided, data-driven methods, particularly machine learning (ML), are attractive methods for calculating spectro-photometric distances. \citet{leung_deep_2019} determined the spectro-photometric distances for stars observed by the Apache Point Observatory Galactic Evolution Experiment (APOGEE; \citealp{majewski_apache_2017}) and provided a flexible model to calibrate parallax zero-point biases in Gaia Data Release 2 (DR2).  
\citet{stone-martinez_spectroscopic_2024} used a simple neural network to drive the absolute magnitude from stellar parameters and provided a value-added catalog (VAC) with 733,901 spectra.

Mapping the stellar parameters derived from spectra to absolute magnitude, the ML technologies have been successfully applied. 
\citet{zhang_deriving_2020} proposed the Stellar LAbel Machine (\texttt{SLAM}), which is a data-driven method based on the support vector regression (SVR), shows high performance in deriving stellar parameters from the Large Sky Area Multi-Object Fiber Spectroscopic Telescope (LAMOST; \citealp{luo_first_2015, zhao_lamost_2012, cui_large_2012}) low-resolution spectra.
\citet{fabbro2018} introduced \texttt{StarNet}, leveraging APOGEE observed spectra alongside corresponding ASSET synthetic data. 
\citet{zhang_938720_2019} employed \texttt{StarNet} to analyze giants using low-resolution spectra 
from LAMOST. 
\citet{leung_deep_2019} presented \texttt{astroNN}, an open-source Python tool with flexible neural network features. 
\citet{wang_analysis_2019} introduced generative spectrum networks (\texttt{GSN}) for synthesizing spectra, 
and integrated a Bayesian framework for inferring stellar atmospheric parameters from low-resolution LAMOST spectroscopy. 
Following this, \citet{wang_spcanet_2020} developed \texttt{SPCANet}, which was specifically designed for medium-resolution LAMOST spectroscopy and produced 13 high-precision chemical abundances. 
\citet{ting_payne_2019} presented The \texttt{Payne}, an approach enabling the precise interpolation and prediction of spectra within high-dimensional label spaces. Building upon The \texttt{Payne}, 
\citet{xiang_stellar_2022} introduced \texttt{HotPayne}, tailored for 330,000 hot stars using LAMOST low-resolution spectra. 
By predicting projected rotation velocity and micro-turbulence velocity, \texttt{HotPayne} expands its utility. 
\texttt{Cycle-StarNet} was developed by \citet{obriain_cycle-starnet_2021} using a hybrid generative domain-adaptation approach to convert simulated star spectra into their real-world equivalents. 
By effectively applying this methodology to APOGEE, the discrepancy between theoretical predictions and observational data was reduced. Expanding the scope, \citet{wang_stellar_2023} implemented \texttt{Cycle-StarNet} on LAMOST medium-resolution spectra, employing the L-BFGS algorithm to optimize the fitting of synthetic spectra to estimate stellar parameters and 11 chemical abundances in a staggering dataset of 1.38 million FGKM-type stars. 

In this study we focus on solar-like stars in LAMOST DR9. We found that at least 521,424 of these stars (representing 13\% of the LAMOST solar-like sample) lack precise distance measurements, with uncertainties higher than 20\% or even without any distance estimations when checked with Gaia. Similar to the approaches of \citet{leung_deep_2019}, we use a convolutional neural network (CNN) to predict the absolute magnitudes, colors, and stellar parameters directly from the spectra. With the predicted absolute magnitudes, we can subsequently calculate the distances using the observed apparent magnitudes and extinction values. This approach provides a significant supplement to distance measurements for solar-like stars with inadequate astrometric and photometric data.

This paper is organized as follows. Section \ref{sec:data} describes the data used in this study, including LAMOST low-resolution spectra, stellar atmosphere parameters from APOGEE, and astrometric data from Gaia. In Sect. \ref{sec:method} the CNN model architecture, model training, model performance, and model comparison are detailed. In Sect. \ref{sec:results} the trained model is applied to 521,424 solar-like stars from LAMOST DR9 to derive distances and stellar parameters. The validation and analysis of the results, along with the metallicity gradients of the Milky Way calculated from a subsample, are presented. Finally, in Sect. \ref{sec:discussion and conclusion} we discuss the feature importance and present the conclusions of this work.

\section{Data}
\label{sec:data} 
For this work we employed low-resolution spectra from LAMOST DR9, astrometric and photometric data from Gaia DR3, and stellar atmospheric parameters from APOGEE DR17. The stellar parameters (\teff, \logg, \feh) were obtained from the APOGEE Stellar Parameters and Abundances Pipeline (ASPCAP), and the absolute magnitudes were derived using geometric distances from \citet{bailer-jones_estimating_2021}.

\subsection{Data selection and calculation}
\label{sec:data selection} 
High-quality low-resolution spectra of solar-like stars are selected from the \texttt{LAMOST LRS Stellar Parameter Catalog of A, F, G and K Stars} of LAMOST DR9 v1.1 (hereafter \texttt{LAMOST LRS AFGK Catalog}),\footnote{\url{https://www.lamost.org/dr9/}} which contains 7,060,679 spectra with determined stellar parameters (\teff, \logg, \feh) by the official LAMOST Stellar Parameter Pipeline (LASP). A total of 4,067,525 solar-like stellar spectra were selected according to the criteria in Table \ref{tab:paramrange}, which is similar to the criteria used by \citet{Zhang2022}.
\begin{table}
	\centering
	\caption{Selection criteria for solar-like stars.}
	\label{tab:paramrange}
	\begin{tabular}{lc}
		\hline \hline
        \noalign{\smallskip}  
		parameters & condition \\
        \noalign{\smallskip}
		\hline
        \noalign{\smallskip}
		\teff & 4800 $\leq$ $T\rm_{eff}$ $\leq$ 6800K \\

		\logg & $\logg \leq 5.98-0.00035 \times \teff$ \\

		[Fe/H] & -1.0 $<$ [Fe/H] $<$ 1.0  \\
        \noalign{\smallskip}
		\hline \hline
	\end{tabular}
\end{table}

\begin{figure}
	\centering
	\includegraphics[width=\columnwidth]{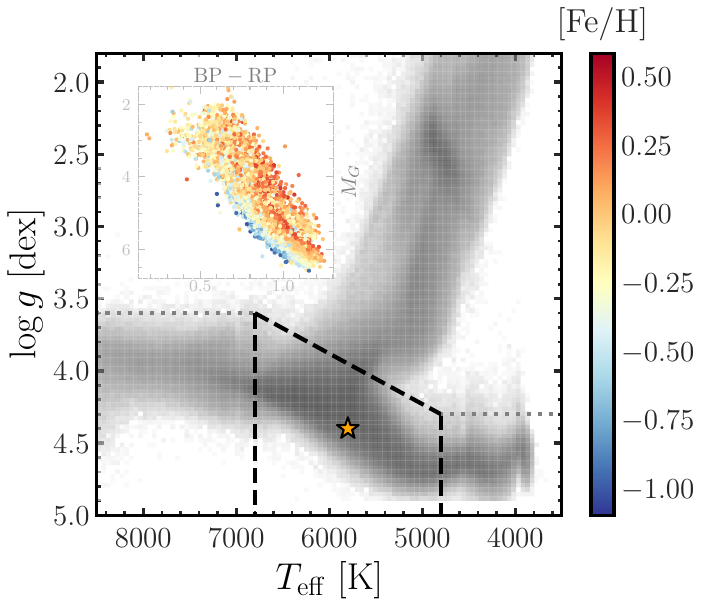}
    \caption{$\teff-\logg$ Kiel diagram of LAMOST DR9 AFGK-type stars. The black dashed line represents the empirical formula proposed by \citet{Zhang2022}, with an effective temperature range of $4800-6800$ K; stars within this area are defined as solar-like stars in this work. For reference, the Sun ($\teff \sim 5800$ K, $\logg \sim 4.4$ dex) is denoted by an orange star. The inset color-magnitude diagram shows the distribution of the training data, which is a subset of stars within the black dashed area, color-coded by \feh.}
    \label{fig:kiel}
\end{figure}

Solar-like stars of the late F-type, G-type, and early K-type are included in this study. The effective temperature range for these stars ranges from $4800$ K to $6800$ K, which encompasses the effective temperature of the Sun ($\sim$ 5800K) with a fluctuation of $\pm$ 1000K. The surface gravity \logg \space is constrained by the empirical formula proposed by \citet{Zhang2022} in the \teff-\logg \space Kiel diagram, as shown in Fig. \ref{fig:kiel} and Table \ref{tab:paramrange}. This empirical formula  distinguishes between main-sequence stars and giants. The metallicity range is set to $-1.0 < \feh < 1.0$ dex, which is around the metallicity of the Sun ($\feh = 0.0$). 

In the \texttt{LAMOST LRS AFGK catalog}, there are 4,022,857 solar-like stars with Gaia data. To find the most reliable distance for training, we considered three primary sources of distance data: inverse parallax (including raw parallax data and parallax after zero-point correction), distance from the General Stellar Parametrizer from Photometry (GSP-Phot; \citealp{GSP-Phot}), and distance obtained by \citet{bailer-jones_estimating_2021} using the geometric methods. Before calculating the inverse parallax, the zero-point correction for parallax is determined separately for five-parameter and six-parameter astrometric solutions, as explained in \citet{Lindegren2021}. The method for calculating zero points follows the Python code.\footnote{\url{https://gitlab.com/icc-ub/public/gaiadr3\_zeropoint}} 

\begin{figure}
	\centering
	\includegraphics[width=\columnwidth]{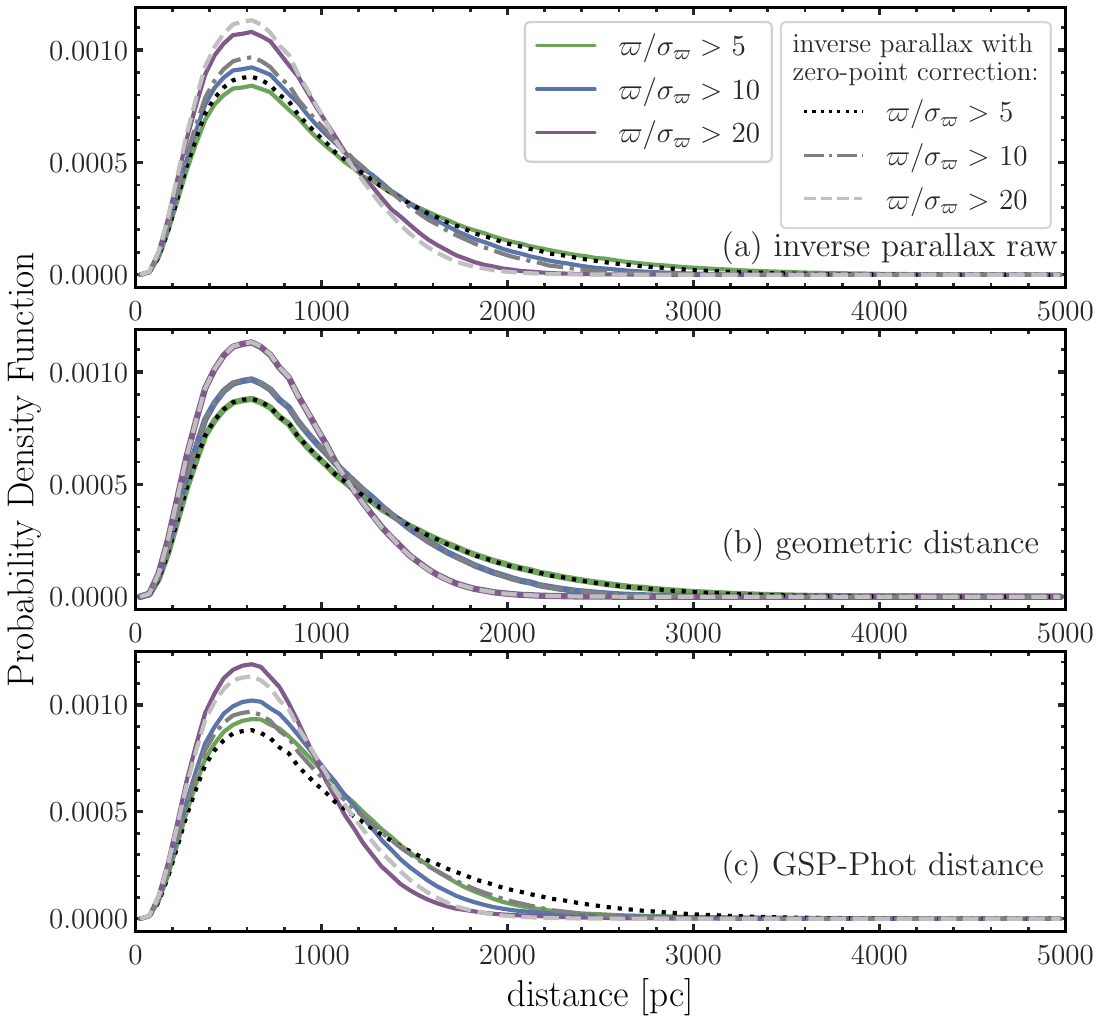}
    \caption{Distributions of distances from different sources in the solar-like star dataset. The green, blue, and purple lines represent the distributions with $\varpi / \sigma_{\varpi} > 5$, $\varpi / \sigma_{\varpi} > 10$, and $\varpi / \sigma_{\varpi} > 20$, respectively. Panels (a), (b), and (c) show the distributions of raw inverse parallax, geometric distances, and GSP-Phot distances. The distributions of inverse parallax with zero-point correction for different error cuts are shown as black dotted, gray dot-dashed, and silver dashed lines, respectively. For comparison, these distributions are shown in all three panels.}
    \label{fig:distance-pdf}
\end{figure}

According to the distribution of distances shown in Fig. \ref{fig:distance-pdf}, as $\varpi / \sigma_{\varpi}$ increases, the amount of distance data decreases, especially for distances greater than 1 kpc. 
Compared to the corrected inverse parallax, the raw inverse parallax has a larger dispersion and relatively overestimates distances in panel (a). In panel (b), the geometric distance has a similar distribution to the corrected inverse parallax. In panel (c), the GSP-Phot distance has a smaller dispersion at close distances and tends to underestimate the distance.

\begin{figure}
	\centering
	\includegraphics[width=\columnwidth]{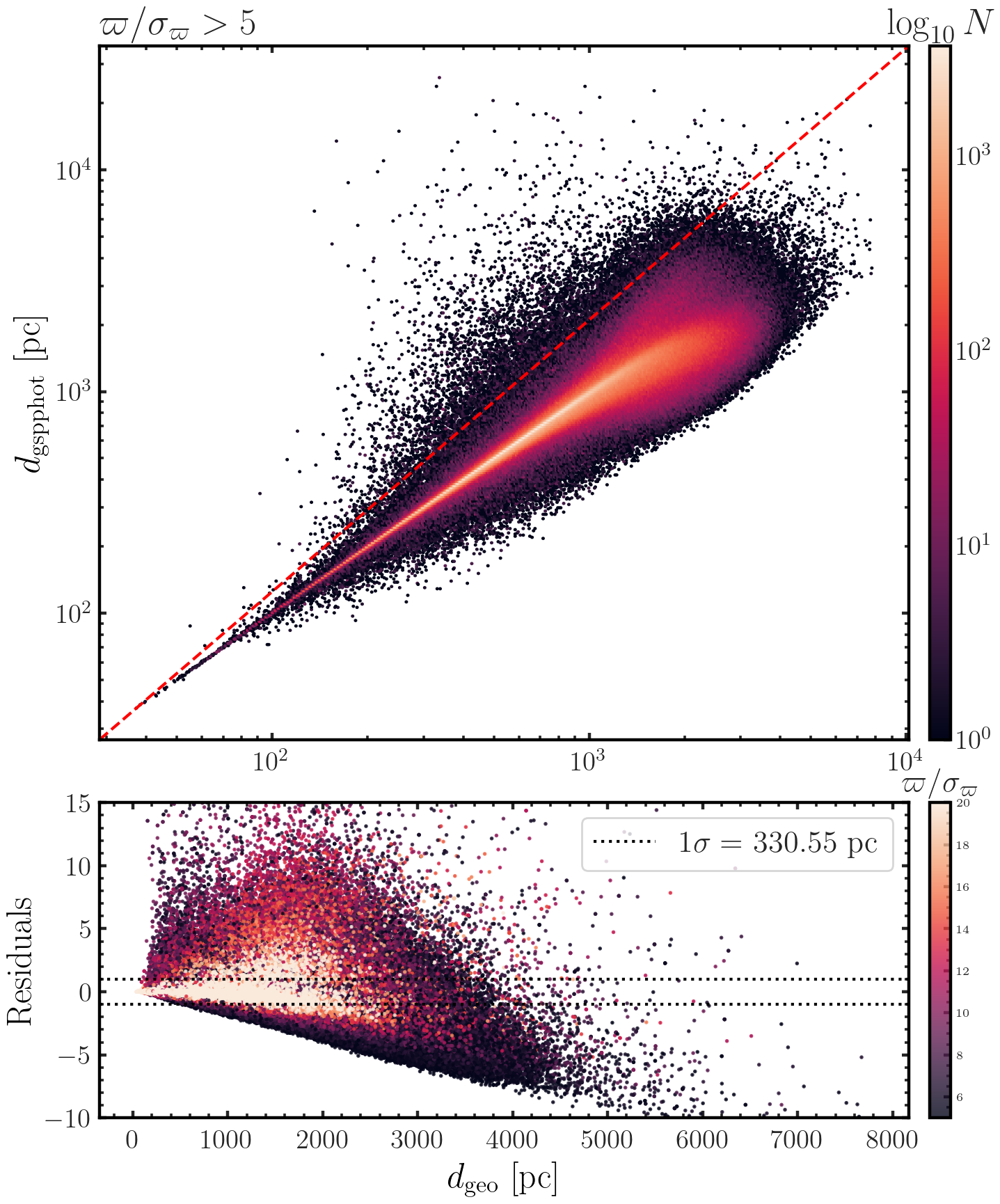}
	\caption{Comparison of distance from GSP-Phot and geometric in solar-like dataset with $\varpi / \sigma_{\varpi} > 5$. The upper panel illustrates how GSP-Phot significantly underestimates distance when compared to distance from geometric, and how the underestimation worsens with increasing distance. In the upper panel, the red dotted line is the 1:1 diagonal, and the color-coding reflects the number of log forms. The lower panel displays the variation of the residuals with distance, the color-coding represents the $\varpi / \sigma_{\varpi}$ values. The variance of the residuals gets larger as the distance grows, but the situation is less severe for high  $\varpi / \sigma_{\varpi}$.}
    \label{fig:distance-mix}
\end{figure}

In Fig. \ref{fig:distance-mix} GSP-Phot distances and geometric distances are shown for $\varpi / \sigma_{\varpi} > 5$. GSP-Phot distances are significantly underestimated when compared to geometric distances, and this underestimation worsens at larger distances. However, for larger $\varpi / \sigma_{\varpi}$, indicating higher data quality, the underestimation is mitigated with increasing distance. 
In this study, geometric distances, which are derived purely through a Bayesian approach that excludes the influence of photometric data, are chosen to determine absolute magnitudes.

We calculated the absolute magnitude from the known apparent magnitude \texttt{g\_mean\_mag} in the Gaia DR3 catalog and the geometric distance $d_{\rm{geo}}$ (\texttt{r\_med\_geo}) from \citet{bailer-jones_estimating_2021}:

\begin{equation}
\label{eq:mag-dis}
    M_{\rm{G}} = m_{G} - 5 \log{d_{\rm{geo}}} + 5 - A_{\rm{G}},
\end{equation}
Here $M_{\rm{G}}$ is the calculated absolute magnitude, which is used as the label for the training process and  $A_{\rm{G}}$ (\texttt{ag\_gspphot})  is the extinction in G band from GSP-Phot Aeneas best library using BP/RP spectra. Additionally, we use the extinction information from GSP-Phot (\texttt{ebpminrp\_gspphot}) to determine \bprp \space of the color index $(\rm{{BP} - {RP}})$ (\texttt{bp\_rp}).

Finally, a dataset containing 18,573 stars with stellar atmosphere parameters provided by APOGEE is used for training in this study. Table \ref{tab:dataset} details the step-by-step filtering process of the training data, including the filtering criteria and the remaining number of spectra at each stage. The color-magnitude diagram of the training data, color-coded by metallicity from APOGEE, is shown in the inset of Fig. \ref{fig:kiel}. Additionally, the number density distribution of each parameter is presented in Fig. \ref{fig:param-pdf}; further details on how the reference and test sets are distinguished can be found in Sect. \ref{sec:training}. 
\setlength{\tabcolsep}{1.5pt}
\begin{table}
	\centering
	\caption{Cuts applied for the training dataset.}
	\label{tab:dataset}
	\begin{tabular}{lc}
		\hline \hline
        \noalign{\smallskip}
		Cut & Number of spectra \\
        \noalign{\smallskip}
		\hline
        \noalign{\smallskip}
		\texttt{LAMOST LRS AFGK Catalog} & 7,060,679 \\

		  solar-like stars according to LASP & 4,067,525 \\

		  APOGEE DR17 cross-match within 1 arcsec & 54,971  \\
		  
		  solar-like in APOGEE & 47,557  \\
		  
		  Gaia DR3 cross-match within 3 arcsec & 44,593 \\
		  
		  \texttt{parallax\_over\_error} $> 5$ & 39,416 \\
		  
		  $\rm{S/N_g} > 10 $ and \texttt{fibermask} $= 0$  & 18,573 \\

		\noalign{\smallskip}
		\hline \hline
	\end{tabular}
\end{table}

\subsection{Pre-processing of the input data}
\label{sec:preprocess}
The radial velocity of the object causes the spectral lines to undergo the Doppler effect. Hence, for the input spectrum, we first needed to convert the wavelengths to the rest frame.

Second, to ensure alignment, we interpolated the raw spectra to a uniform wavelength range. Typically, both ends of the spectrum are noisy and contain less information about the spectral features. Therefore, we interpolated the entire sample set to a wavelength range of 3925--8800\AA, using 1\AA \space as the sampling point, resulting in a total of 4875 data points for spectral flux data.

Finally, each stellar spectrum in the training set was normalized by dividing it with its pseudocontinuum. The pseudocontinuum was obtained by fitting a polynomial to the spectrum. For this work, we found that despite \texttt{FIBERMAS=0}, there are still bad pixels that were not completely removed, leading to incorrect spectral lines and extremely high pixel values that affected the model prediction. To address this, we implemented two additional steps.  First, the original spectrum was smoothed using the Savitzky-Golay filter, and then the smoothed spectrum was normalized. Using this approach, the normalized spectra outperformed those provided in the official LAMOST pipelines. Second, the pixels were thresholded in the normalized spectrum by simply setting to zero those above 1.2 and below 0.1. This straightforward threshold processing is effective because the anomalous pixels appear at random wavelengths. It is believed that as long as their values are not too far from the normalized flux range ($0-1$), individual pixels do not significantly affect the prediction ability of the neural network. Two examples illustrating these preprocessing steps are shown in Fig. \ref{fig:preprocess}.

\section{Method}
\label{sec:method}
\subsection{The CNN architecture}
\label{sec:cnn}
A neural network has three components: an input layer, some hidden layers, and an output layer. Given a training set of input spectra with known labels (\teff, \logg, \feh, \absm, \bprp), a neural network model can generate a function that maps the input spectra to the output parameters. This nonlinear capability is attributed to the activation function, which determines the significance of each neuron in the network. The trained model can then be used to predict the physical properties of spectra with unknown stellar parameters in another dataset. 
 
\begin{figure*}
	\centering
	\includegraphics*[width=\textwidth]{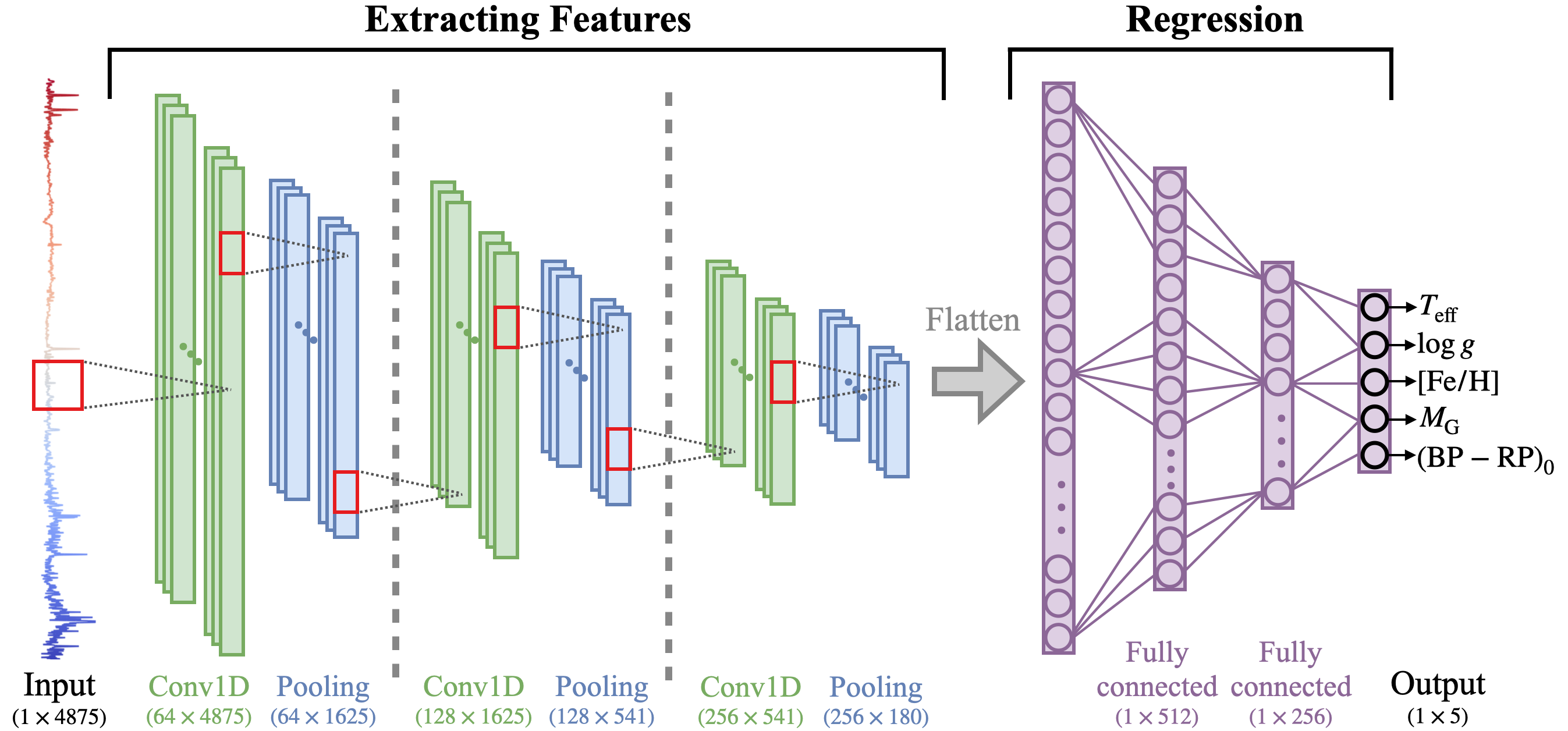}
	\caption{Ten-layer architecture of the CNN model used in this study. The first layer is solely the input data, followed by three convolutional blocks, each composed of a convolutional layer and a max pooling layer, both with a size of three units. This is followed by three fully connected layers with 512, 256, and 5 nodes, respectively, where the final layer serves as the output layer.}
	\label{fig:cnn-architecture}
\end{figure*}

In this work, we propose a one-dimensional convolutional neural network (CNN) regression model. The architecture of the CNN model is shown in Fig. \ref{fig:cnn-architecture}. It consists of three convolutional blocks for feature extraction and two fully connected layers for regression. The activation function for both the convolutional and fully connected layers is Leaky ReLU. The output layer uses a linear activation function and has five neurons, each corresponding to a stellar parameter.

We tested different network architectures specifically for our dataset, including various combinations of convolutional blocks and fully connected layers, similar to Fig. 3 in \citet{zhang_938720_2019}. We found that the architecture with three convolutional blocks and two fully connected layers is well-suited for our dataset. It is important to note that for different tasks and datasets, the model must be modified or fine-tuned.

\subsection{Model training}
\label{sec:training}
The training process was carried out using the dataset described in Sect. \ref{sec:data selection}, which was divided into reference and test sets in an 8:2 ratio. A total of 14,858 stars were used for training and cross-validation in the reference set, while 3,715 stars were used in the test set to evaluate the performance of the trained models. Before training, the input pseudo-continuous spectra were already normalized, while the output labels were standardized using Z-score standardization.

To ensure robust model performance, we employed five-fold cross-validation, dividing the reference set into five non-overlapping subsets. One subset served as the validation set, and the remaining four served as the training set. This strategy ensured the efficient use of data and the proper examination of the model's generalization ability by utilizing the entire dataset for both training and validation.

For the model optimization we used the mean squared error (MSE) loss function, defined in Eq. \ref{eq:mse}, to measure the difference between the target and predicted parameters:

\begin{align}
	\label{eq:mse}
	\rm{MSE} = \frac{1}{n} \sum_{i=1}^{n}(y_i - \hat{y_i})^2
\end{align}
Here $n$ is the number of samples, $y_i$ is the true value, and $\hat{y_i}$ is the model's predicted value. The Adam optimizer \citep{adam} was employed for gradient descent, with an initial learning rate of 0.0001. This rate automatically decreases using the adaptive learning rate tuning method, ReduceOnPlateau \citep{reduceOnPlateau}, which halves the learning rate if the MSE value does not improve after five epochs. This approach allows the model to find a more precise solution and avoids premature convergence to a local optimum.

The CNN model was trained using the NN library \texttt{Keras} \citep{chollet2015keras}, which provides a high-level API for the \texttt{TensorFlow} \citep{tensorflow2015-whitepaper} machine intelligence package. The train and test split, standardization, and $k$-fold cross-validation were performed using the \texttt{scikit-learn} library \citep{scikit-learn}.

\subsection{Model performance and error analysis}
\label{sec:performance}
\begin{figure*}
	\centering
	\includegraphics[width=\textwidth]{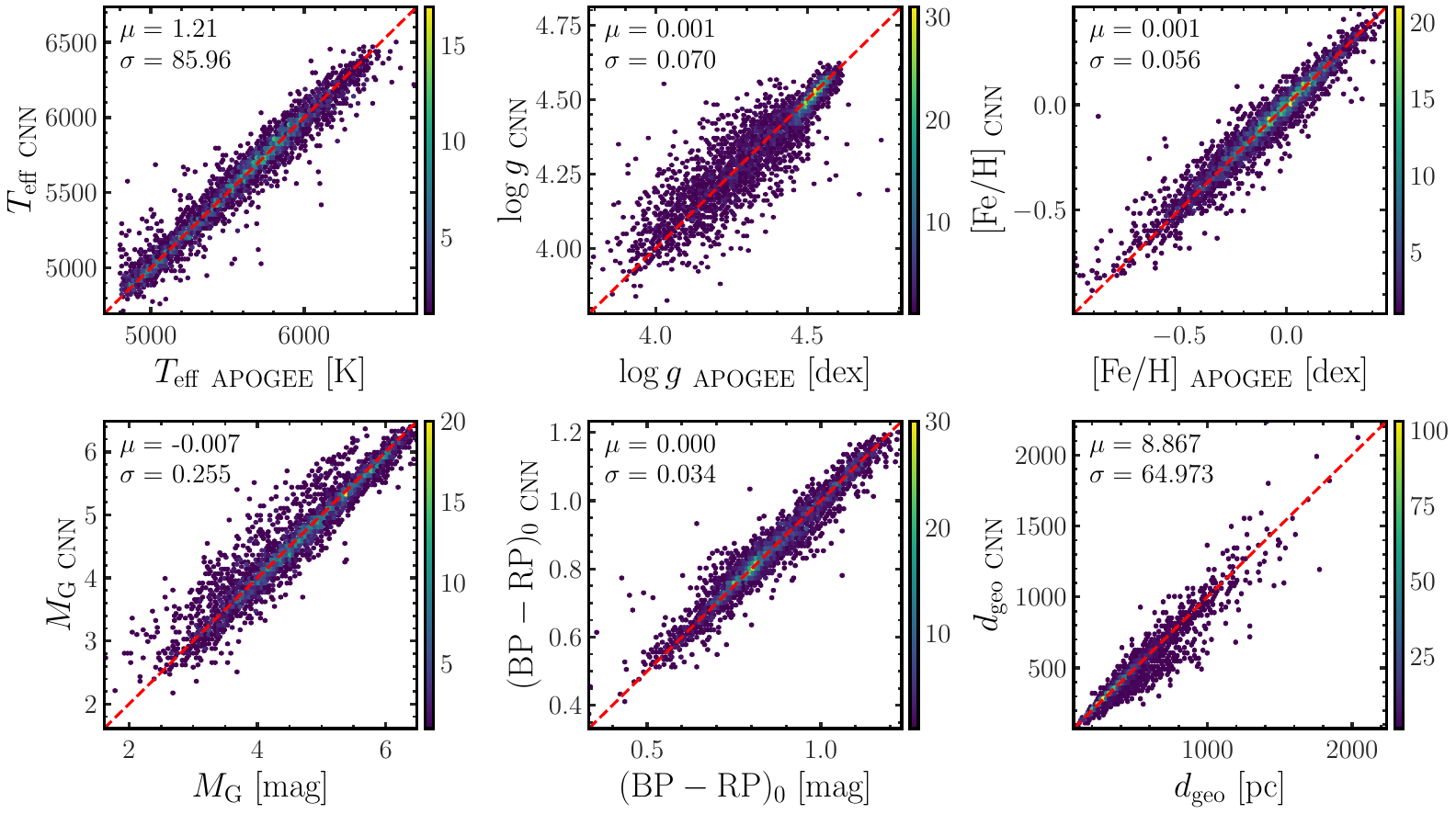}
    \caption{Comparison of the predictions and ASPCAP parameters on a test set of 3,715 stars. The bottom right panel compares the distances calculated from the predicted absolute magnitudes \absm \space and the geometric distances. The red dashed line in each panel represents the 1:1 line.}
    \label{fig:test_result}
\end{figure*}

The performance of the CNN model was evaluated on the test set, as shown in Fig. \ref{fig:test_result}. The results indicate that the CNN model's predictions for \teff, \logg, and \feh \space closely align with the APOGEE parameters, while \absm \space and \bprp \space closely match the Gaia parameters. There is no significant bias for these parameters. The scatter is 86 K for \teff, 0.07 dex for \logg, 0.06 dex for \feh, 0.25 mag for \absm, and 0.03 mag for \bprp.
To calculate the distance, we assumed $M_{\rm{true}} \sim \mathcal{N}(\mu_{M},\sigma_{M})$, where $\mu_{M}$ is the \absm \space predicted by CNN, and $\sigma_{M}$ is the scatter between the labeled values and the predicted values, as defined in Eq. \ref{eq:p1}. The predicted absolute magnitudes \absm \space were then used to estimate the distance distribution, as shown in Eq. \ref{eq:p2}: 
\begin{align}
    P_{M}(M\mid \mu_{M},\sigma_{M}) &=\dfrac{1}{\sqrt{2\pi }\sigma_{M}}\cdot \exp[-\dfrac{\left( M-\mu_{M}\right) ^{2}}{2\sigma_{M}^{2}}] \label{eq:p1} \\
    h(d) &= m-5\log d+5-A_{\rm{G}} \label{eq:h} \\
    P_{d}(d\mid m,A_{\rm{G}},\mu_{M},\sigma_{M}) &=\left| h(d) '\right| \cdot P_{M}\left[h(d)\right] \label{eq:p2}
\end{align}
These estimated distances are illustrated in the lower right panel of Fig. \ref{fig:test_result}, where the predicted distances are compared to the geometric distances, with a scatter of 65 pc. We also calculated the fractional error, defined as the difference between the predicted and geometric distances divided by the geometric distances. The median fractional distance error is 3.7\% with a standard deviation of 8.2\%, as shown in the fourth panel in Fig. \ref{fig:resid_dist}. It should be noted that our dataset contains some contamination from unresolved binaries. These stars tend to have higher residuals because the model predicts a dimmer luminosity for them, leading to a slight bimodal distribution in the fractional distance error. However, since our focus is on studying solar-like stars, and these potential unresolved binaries are fairly uniformly distributed across the parameter space, their influence on the model is naturally averaged out across each type of spectrum during the training process. Moreover, as shown in Fig. 2 of \citet{El-Badry2018}, for LAMOST-like spectra with $\rm{S/N=30}$, the typical systematic error caused by unresolved binaries is approximately 100 K in \teff, while the errors in \logg \space and \feh \space are smaller than 0.1 dex and 0.05 dex, respectively. This degree of systematic error is acceptable for most scientific objectives.

\begin{figure*}
	\centering
	\includegraphics[width=\textwidth]{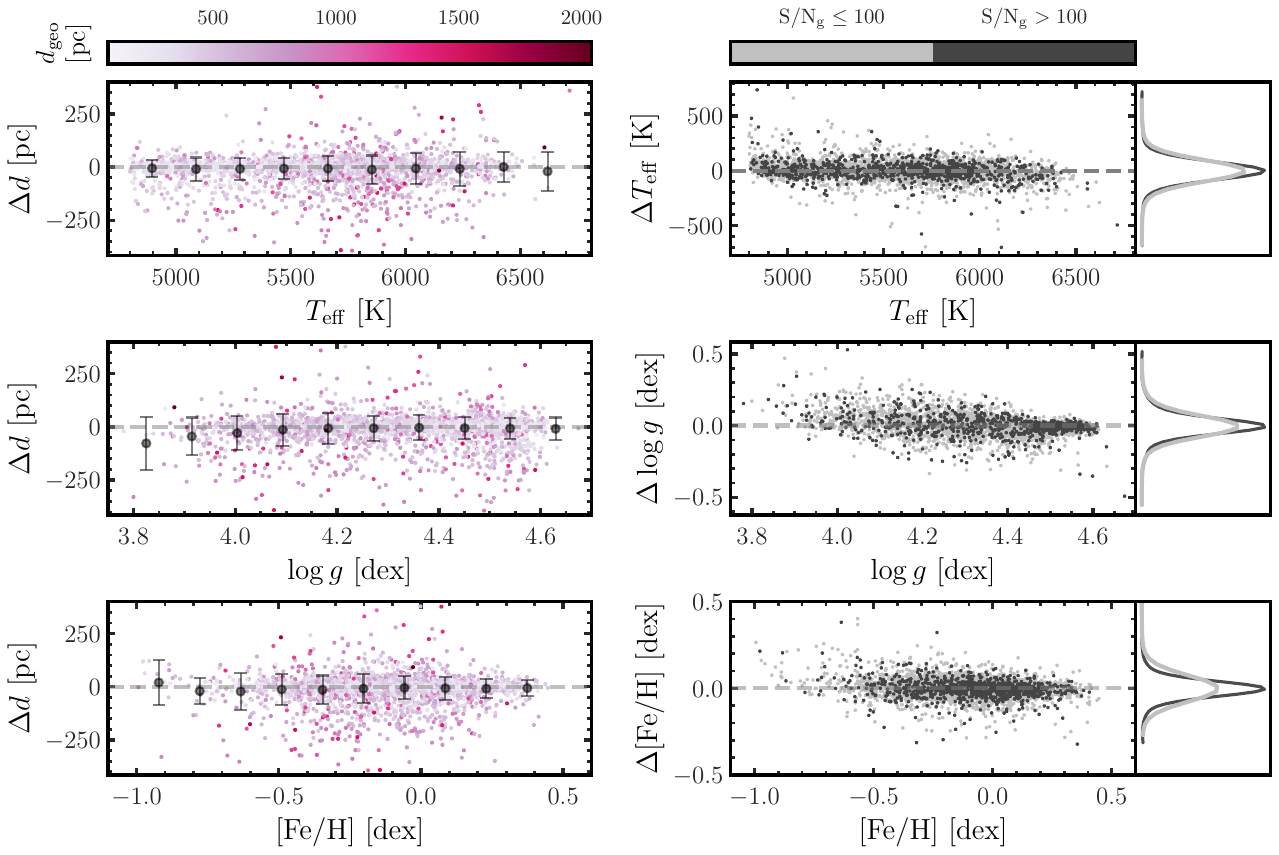}
    \caption{Fractional distance errors and residuals of CNN predictions compared to labeled values from ASPCAP parameters, shown as a function of stellar properties. The residuals are calculated as the predicted values minus the labeled ones. The left three panels show the fractional distance error vs three ASPCAP stellar parameters, color-coded by geometric distance. The gray dots with error bars represent the mean and standard deviation of the error in each bin. The right three panels show the residuals of \teff, \logg, and \feh \space vs the ASPCAP values, color-coded by $\rm{S/N_{g}}$ (gray for spectra with $\rm{S/N_{g}} \leq 100$ and black for spectra with $\rm{S/N_{g}} > 100$). The projected residual distributions are shown on the right.}
    \label{fig:resid_param}
\end{figure*}

\begin{figure*}
	\centering
	\includegraphics[width=\textwidth]{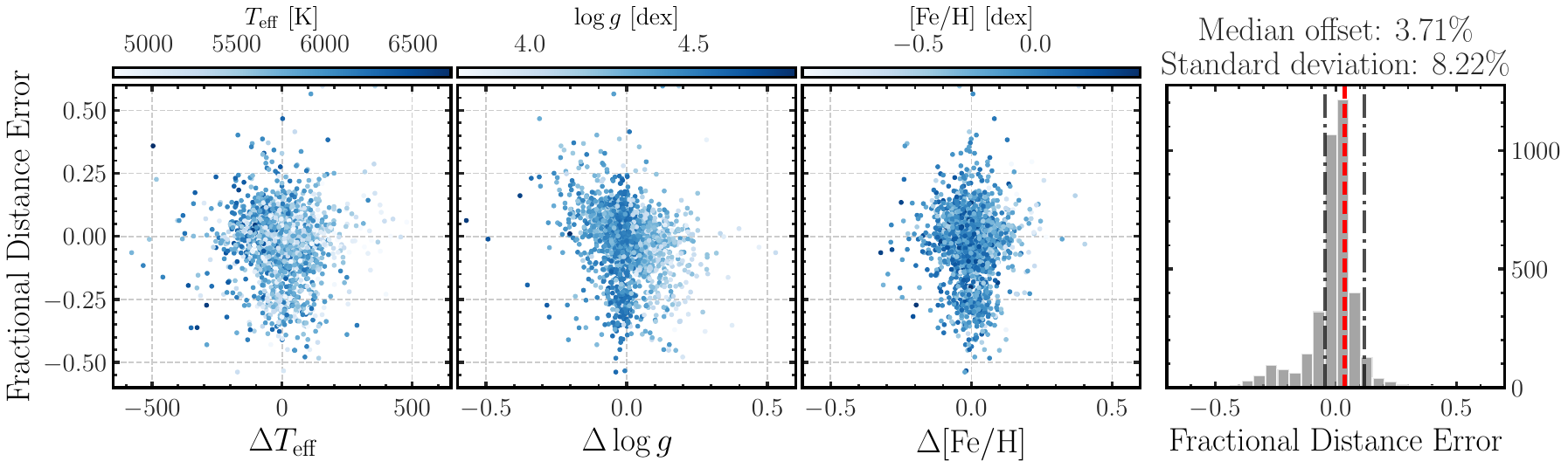}
    \caption{Fractional distance error compared to the residuals of stellar parameters, and the distribution of fractional distance error. The left three panels show the fractional distance error vs the residuals of \teff, \logg, and \feh, color-coded by the labeled values from ASPCAP. The last panel shows the distribution of fractional distance error, with a median of 3.7\% (red dashed line) and a standard deviation of 8.2\% (black dash-dotted line).}
    \label{fig:resid_dist}
\end{figure*}

To further evaluate, we compared the fractional distance error and residual of the CNN predictions to the labeled values with \teff, \logg, and \feh \space separately, as shown in Fig. \ref{fig:resid_param}.
In the left three panels, the mean fractional distance errors (gray dots) do not exhibit significant deviations from \teff. However, the standard deviation of fractional error (represented by the error bars on the gray dots) shows a slight increase at higher \teff, lower \logg, and lower \feh. At lower \logg \space($\logg < 4.0$) and \feh \space($\feh < -0.5$), our model shows larger deviations from the Gaia distances. These deviations likely arise because the quality of the model predictions depends on the number of stars in the training set that span the parameter space of the test set. Consequently, we suggest that the deviations are due to an insufficient number of stars in our reference set, which is a common issue in data-driven models. Additionally, the precision of predicted distances improves with smaller distances due to Gaia's higher accuracy in these ranges. 

The right three panels of Fig. \ref{fig:resid_param} show the residuals of \teff, \logg, and \feh, respectively. The model predictions show excellent agreement with the ASPCAP results for the high $\rm{S/N_g}$ spectra. However, larger scatters are observed between CNN and ASPCAP for the lower $\rm{S/N_g}$ spectra. From the projected residual distributions, \feh \space is the most sensitive parameter to the signal-to-noise ratio, while \teff \space and \logg \space are less sensitive. The metallicity depends on the strength of the spectral lines, which are more affected by the signal-to-noise ratio, whereas \teff \space and \logg \space rely on the continuum shape and are less affected. Meanwhile, for lower effective temperatures ($\teff < 5000$ K), lower surface gravity ($\logg < 4.1$), and lower metallicity ($\feh < -0.5$), the CNN model tends to predict higher values compared to ASPCAP. This discrepancy is likely due to the nature of the spectra themselves, where weaker spectral features would make it challenging for the CNN model to identify the most significant features during training. Similarly, the parameters determined by ASPCAP have larger intrinsic uncertainties, which in turn affect the model performance.

In this work we find no strong evidence that the derived distance errors are affected by surface gravity, as reported in previous studies \citep{estimate_jeffrey_2015, leung_simultaneous_2019,stone-martinez_spectroscopic_2024}; this is due to the limited surface gravity range of our dataset, which is mostly concentrated around $3.5 - 4.5$ dex for dwarf stars. In the left three panels of Fig. \ref{fig:resid_dist}, the fractional distance error is compared to the errors in \teff, \logg, and \feh. There is no significant dependence of distance error on the uncertainties in these three stellar parameters. The second panel shows a slight correlation between $\Delta \logg$ and \logg: the model tends to overestimate \logg \space at lower values and underestimate it at higher values, which is also reflected in the second panel on the right in Fig. \ref{fig:resid_param}. The last panel shows the distribution of fractional distance error, with a median of 3.7\% and a standard deviation of 8.2\%, which demonstrate that this CNN model has a profound ability for distance estimations.

We also compared our CNN model with other ML algorithms, such as LightGBM boosted trees \citep{ke2017lightgbm}, Random Forests, and $k$-Nearest Neighbors. We find that our CNN model is a better choice in both precision and efficiency; the details are described in Appendix \ref{sec:model comparison}.

\section{Results}
\label{sec:results}
\subsection{The solar-like stars from LAMOST DR9}
\label{sec:application sample}
\begin{figure}
	\centering
	\includegraphics[width=0.95\columnwidth]{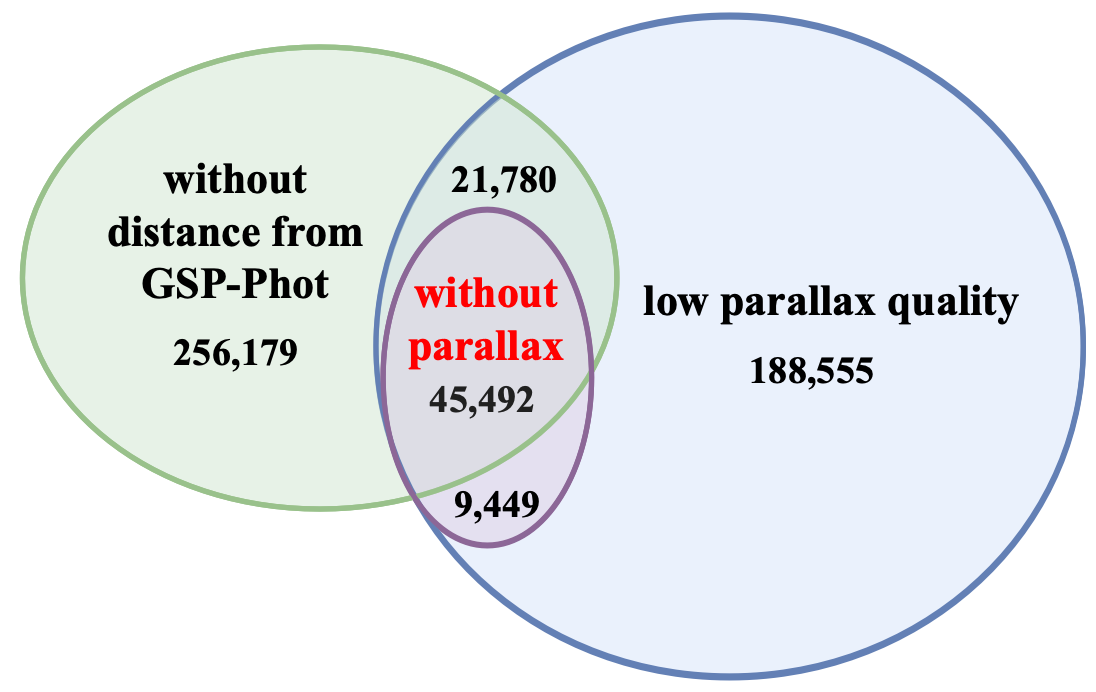}
    \caption{Venn diagram of 512,455 spectra with problematic parallax measurements, including 54,941 spectra without parallax, 265,276 spectra with low parallax quality ($\texttt{parallax\_over\_error} \leq 5$), and 323,451 spectra without GSP-Phot distances. Only 9,449 spectra have distances reported by GSP-Phot in the absence of parallax measurements.}
    \label{fig:venn}
\end{figure}

Based on the  solar-like stars obtained using the second screening condition in Table \ref{tab:dataset}, we selected spectra with ${\rm{S/N_g}} \geq 10$. Among these, we identified spectra with problematic distance measurements: those where Gaia did not provide parallaxes, those where Gaia GSP-Phot did not provide distances, and those where the parallaxes had large errors ($\texttt{parallax\_over\_error} \leq 5$, indicating fractional error larger than 20\%). This selection process resulted in a total of 521,455 spectra, representing 13\% of the LAMOST solar-like sample. More details are shown in the Venn diagram in Fig. \ref{fig:venn}. Even with data lacking parallax values, it can be seen that  GSP-Phot can provide distances in some circumstances; however, only 9,449 spectra meet this condition. Additionally, larger parallax errors lead to less precise distance estimates \citep{gaiachap11}. Fortunately, our model offers a solution for these problematic cases.

\begin{figure}
	\centering
	\includegraphics[width=0.98\columnwidth]{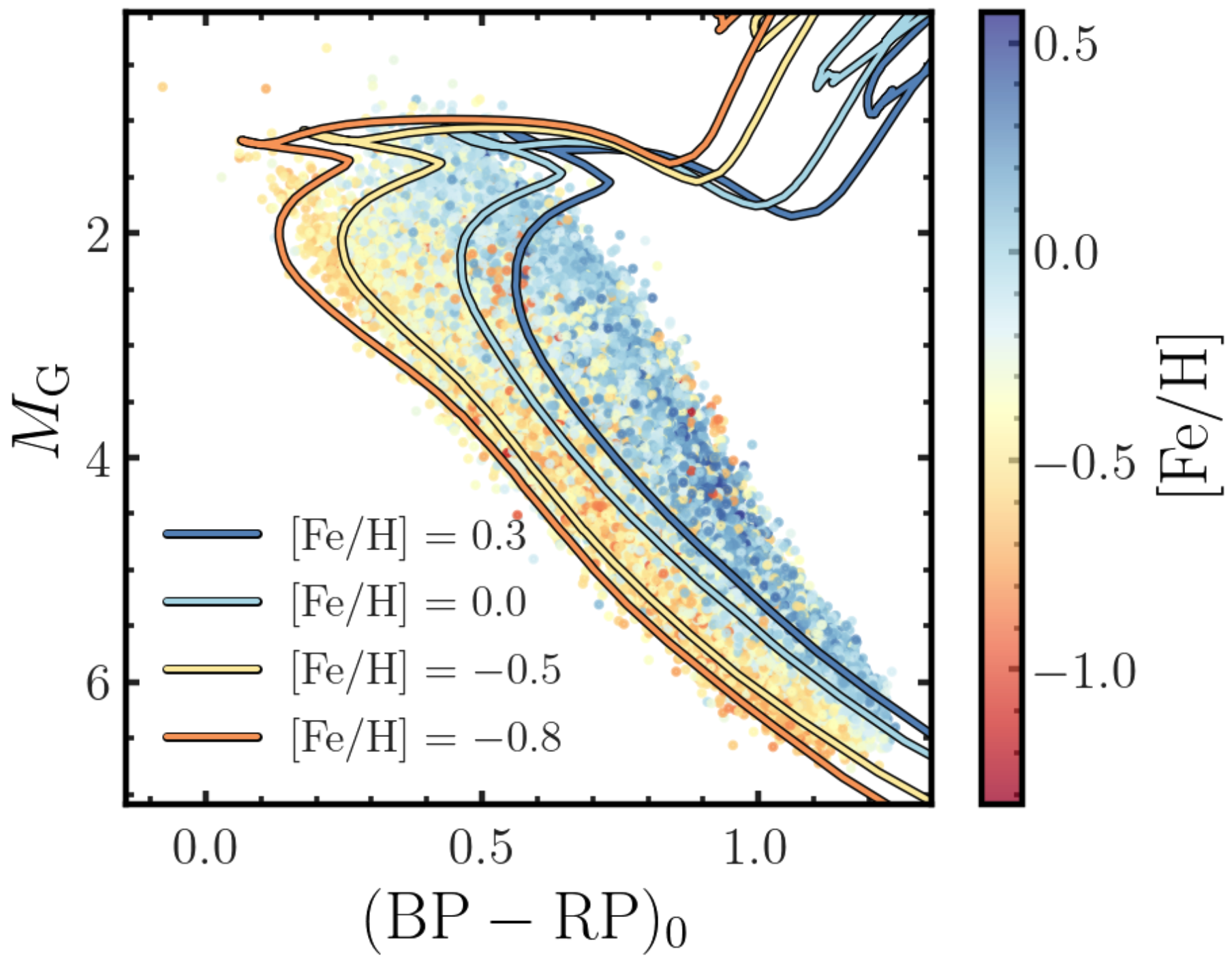}
    \caption{Color-magnitude diagram of 521,424 solar-like stars from LAMOST DR9 with \absm \space and \bprp \space predicted by the CNN model. The colors represent the predicted \feh. Four isochrones with different metallicities ($\feh = -0.8, -0.5, 0.0,$ and $0.3$ dex) at an age of 1.5 Gyr from the PARSEC model are shown in corresponding colors.}
    \label{fig:cmd}
\end{figure}

Using the same preprocessing described in Sect. \ref{sec:preprocess}, we obtained 521,424 normalized spectra spanning wavelengths from 3925\AA \space to 8800\AA. Based on the trained CNN model, we predicted the stellar parameters \teff, \logg, \feh, \absm, and \bprp \space for 521,424 solar-like stars. The results are shown in Fig. \ref{fig:cmd}, which displays a color--magnitude diagram of these stars using the predicted \absm \space and \bprp. The CNN predictions are compared to stellar isochrones to show the expected trend in the parameter relationships. The four lines are the theoretical isochrones from the PAdova and TRieste Stellar Evolution Code (PARSEC) model \citep{bressan2012parsec} with $\feh = -0.8, -0.5, 0.0,$ and $0.3$ dex at an age of 1.5 Gyr, as marked by the corresponding colors. The trend aligns well with the isochrones, though the large scatter is due to the multiple populations in the sample. The results including stellar parameters and distance are saved in a CSV format table and can be found online.\footnote{\url{https://nadc.china-vo.org/res/r101400/}} The column descriptions of the final catalog are given in Table \ref{tab:finalcatalog}.

\subsection{Comparison with other distances}
\label{sec:compdistance}
To ensure the reliability of the distances obtained by our CNN model, we compared them with those from other studies, including \citet{queiroz_starhorse_2018} (hereafter StarHorse), \citet{stone-martinez_spectroscopic_2024} (hereafter DistMass), and \citet{leung_simultaneous_2019} (hereafter astroNN).

StarHorse \citep{santiago_spectro-photometric_2016, queiroz_starhorse_2018, queiroz_bulge_2020} is a Bayesian isochrone-fitting code that derives distances, extinctions, and astrophysical parameters for APOGEE stars, with a typical distance uncertainty of $\sim 5\%$. DistMass \citep{stone-martinez_spectroscopic_2024} is a simple neural network that predicts distances and masses of stars in APOGEE DR17. The distances are trained from stellar parameters using Gaia DR3 distances and literature distances for star clusters, achieving a median fractional distance error of $\sim 10\%$ at higher \logg \space and a standard deviation of $\sim 11\%$. The deep learning model astroNN \citep{leung_deep_2019,leung_simultaneous_2019} uses APOGEE DR17 spectra and Gaia eDR3 data to determine stellar parameters, distances, and other stellar properties. These three catalogs are provided as the  VACs in APOGEE DR17.\footnote{\url{https://www.sdss4.org/dr17/data_access/value-added-catalogs/}}

\begin{figure*}
	\centering
	\includegraphics[width=\textwidth]{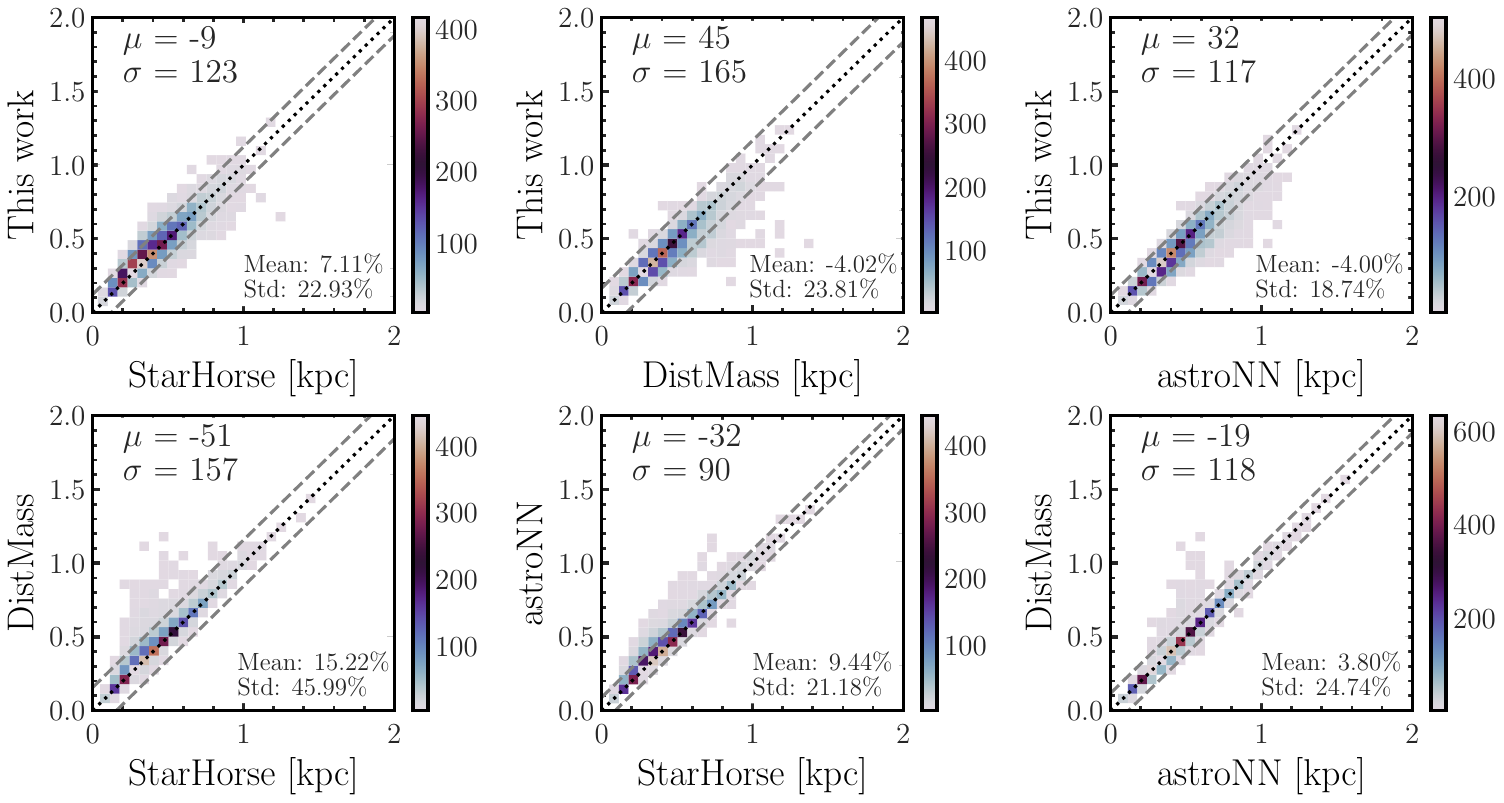}
	\caption{Comparison of estimated distances between this work, StarHorse \citep{queiroz_starhorse_2018,queiroz_bulge_2020}, DistMass \citep{stone-martinez_spectroscopic_2024}, and astroNN \citep{leung_simultaneous_2019}. The top panels show the comparison of 5443, 5240, and 5648 stars in common with StarHorse, DistMass, and astroNN, respectively. The bottom panels show the pairwise comparison between StarHorse, DistMass, and astroNN, with 4401, 4788, and 5242 stars in common, respectively (in the order shown in the figure). The bias $\mu$ (calculated as x-label minus y-label) and scatter $\sigma$ (in units of pc) are indicated in each panel, as well as the corresponding mean and standard deviation of the fractional error. The black dotted lines represent the 1:1 line, and the gray dashed lines indicate the $1\sigma$ deviation in the difference between the compared labels. The color denotes the number density.}
    \label{fig:comp_dist}
\end{figure*}

We cross-matched the  521,424 stars in our results with three catalogs, resulting in 5443, 5240, and 5648 common stars for StarHorse, DistMass, and astroNN, respectively. The top three panels in Fig. \ref{fig:comp_dist} show the comparisons, demonstrating the consistency of the distances obtained by our CNN model with those from other studies. The scatter of the difference between this work and astroNN is 117 pc with an 18.74\% standard deviation of fractional error, which is less than the 123 pc for StarHorse and 165 pc for DistMass. 
The mean fractional error between this work and both DistMass and astroNN is less than 5\%, while for StarHorse it is 7\%. Except for StarHorse, our results appear to be slightly underestimated compared to DistMass and astroNN. This discrepancy may be due to the different training sets and methodologies used in the models.

To provide a comprehensive comparison, we include pairwise comparisons between these studies in the bottom three panels of Fig. \ref{fig:comp_dist}, involving 4401, 4788, and 5242 common stars, respectively. DistMass and astroNN appear to overestimate compared to StarHorse, while DistMass shows higher estimates compared to astroNN. Among these three comparisons, astroNN and DistMass exhibit the best agreement with a 3.8\% mean fractional error, compared to 9.4\% between astroNN and StarHorse, and 15.2\% between DistMass and StarHorse. However, astroNN and StarHorse show the smallest scatter of 90 pc, while it is 118 pc between astroNN and DistMass, and 157 pc between DistMass and StarHorse. The pairwise comparison shows that astroNN and DistMass, both neural network models, exhibit greater consistency with each other than with StarHorse, a Bayesian isochrone-fitting model, reflecting the methodological differences. In summary, our distance estimation agrees well with those of other studies.

\subsection{Comparison with other surveys}
\label{sec:compsurveys}
To ensure the accuracy of the stellar parameters obtained with our CNN model, we compared our results with those from high-resolution observations, specifically the APOGEE DR17 \citep{Abdurro_apogeedr17_2022} and the third data release of the GALactic Archaeology with HERMES surveys (GALAH DR3; \citealp{buder_galah_2021}), two of the most comprehensive spectroscopic surveys.

APOGEE DR17 \citep{Abdurro_apogeedr17_2022} published medium-high-resolution (R $\sim 22,500$) near-infrared spectra of over 650,000 stars from the APOGEE-North and APOGEE-South surveys. The stellar parameters were derived by ASPCAP using MARCS model atmospheres with new spectral grids accounting for non-LTE level populations. GALAH DR3 \citep{buder_galah_2021} includes 768,423 high-resolution (R $\sim 28,000$) optical spectra from 342,682 stars. The parameters were estimated using the Spectroscopy Made Easy (SME) model-driven approach and one-dimensional MARCS model atmospheres. Additionally, GALAH DR3 mitigated spectroscopic degeneracies using astrometry from Gaia DR2 and photometry from 2MASS.

To perform the comparison, we cross-matched our results with APOGEE DR17 and GALAH DR3, identifying 6,040 and 6,329 stars with corresponding stellar parameters. To ensure the reliability of the comparison, we applied \texttt{STARFLAG} $=0$ for the APOGEE spectra; \texttt{flag\_sp} $=0$, \texttt{flag\_fe\_h} $=0$, and \texttt{red\_flag} $=0$ for the GALAH spectra; and \texttt{snrg} $>50$ for the  LAMOST spectra. Moreover, we impose constraints on the errors of \teff, \logg, and \feh \space given by ASPCAP and GALAH, setting them to less than 200 K, 0.2 dex, and 0.2 dex, respectively. Finally, we obtain 2,677 common stars with APOGEE and 2,218 with GALAH.

\begin{figure*}
	\centering
	\includegraphics[width=\textwidth]{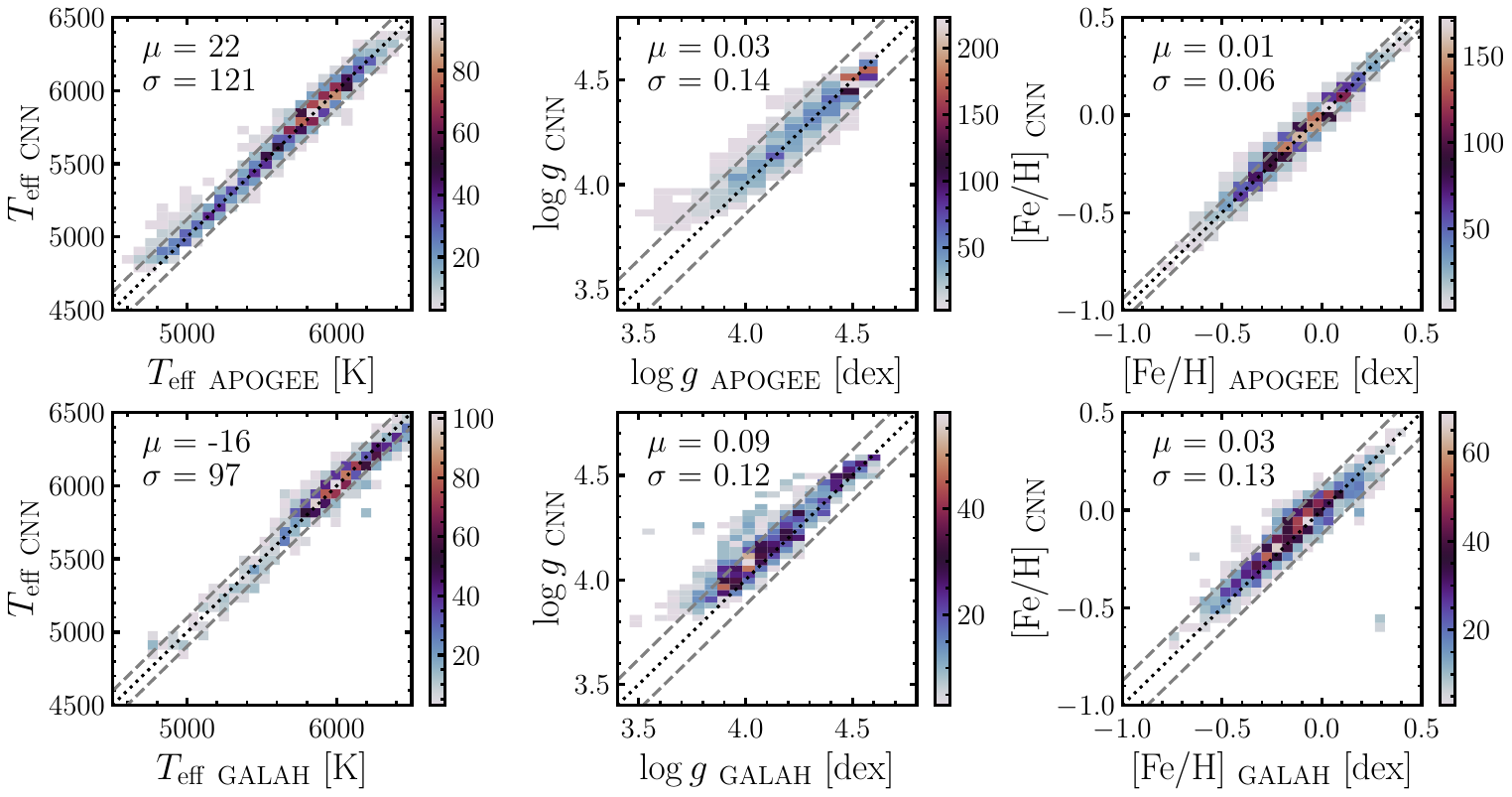}
    \caption{Comparison of the stellar parameters from this work with those from the APOGEE and GALAH surveys. Each column corresponds to a specific parameter: \teff, \logg, or \feh. The top panels represent the comparison between this work and APOGEE DR17 for 2,677 common stars, while the bottom panels show the results with GALAH DR3 for 2,218 stars, both color-coded by number density. The bias $\mu$ (calculated as CNN-label minus survey-label) and scatter $\sigma$, as well as the black dotted line and gray dashed line, are the same as described in Fig. \ref{fig:comp_dist}.}
    \label{fig:comp_param}
\end{figure*}

Fig. \ref{fig:comp_param} shows the differences of \teff, \logg, and \feh \space between this work and the two surveys. In terms of effective temperature, our results are more consistent with those of GALAH, with a scatter of 97 K, while APOGEE shows a scatter of 121 K. This difference may be attributed to the wavelength ranges observed by the surveys, as LAMOST and GALAH both observe in the optical range, while APOGEE in the near-infrared. For surface gravity, the scatter for GALAH is 0.14 dex, which is slightly larger than the 0.12 dex for APOGEE; however, the bias for APOGEE is smaller than that for GALAH. Our results slightly overestimate \logg \space compared to both surveys. Regarding metallicity, this work exhibits more consistency with APOGEE than with GALAH, with APOGEE having a scatter of 0.06 dex compared to 0.13 dex for GALAH. Nonetheless, there is no significant bias for either survey. In summary, our results are in good agreement with both APOGEE and GALAH, demonstrating the reliability of our CNN model.

\subsection{Metallicity gradients of the Milky Way}
\label{sec:metallicity}
Using the  521,424 solar-like stars, we investigated the radial and vertical metallicity gradients to explore the utility of our distance and metallicity measurements for Galactic structure studies. 
Previous studies indicate that the Milky Way's disk shows a negative radial metallicity gradient, with the inner Galaxy (where the galactocentric radius $R$ is less than the solar value) is typically more metal-rich than the outer disk. This gradient ranges from $-0.1 < \Delta\feh/\Delta R < 0.0\ {\rm{dex\ kpc^{-1}}}$ in the Galactic plane. Additionally, the Galactic disk exhibits a negative absolute vertical metallicity gradient, varying in the range $-0.25 < \Delta\feh/\Delta Z < -0.10\ {\rm{dex\ kpc^{-1}}}$, depending on the tracer population and survey volume (\citealp{onal2016local, yan_chemical_2019, gaia_collaboration_metal_2023, wang_spatial_2023, hawkins_chemical_2023, imig_tale_2023}, and references therein). These negative metallicity gradients strongly suggest that the Galactic disk formed in an inside-out manner, with the inner Galaxy forming early and rapidly, followed by the outer Galaxy later on \citep{Frankel_2019}.

\begin{figure}
	\centering
	\includegraphics[width=0.85\columnwidth]{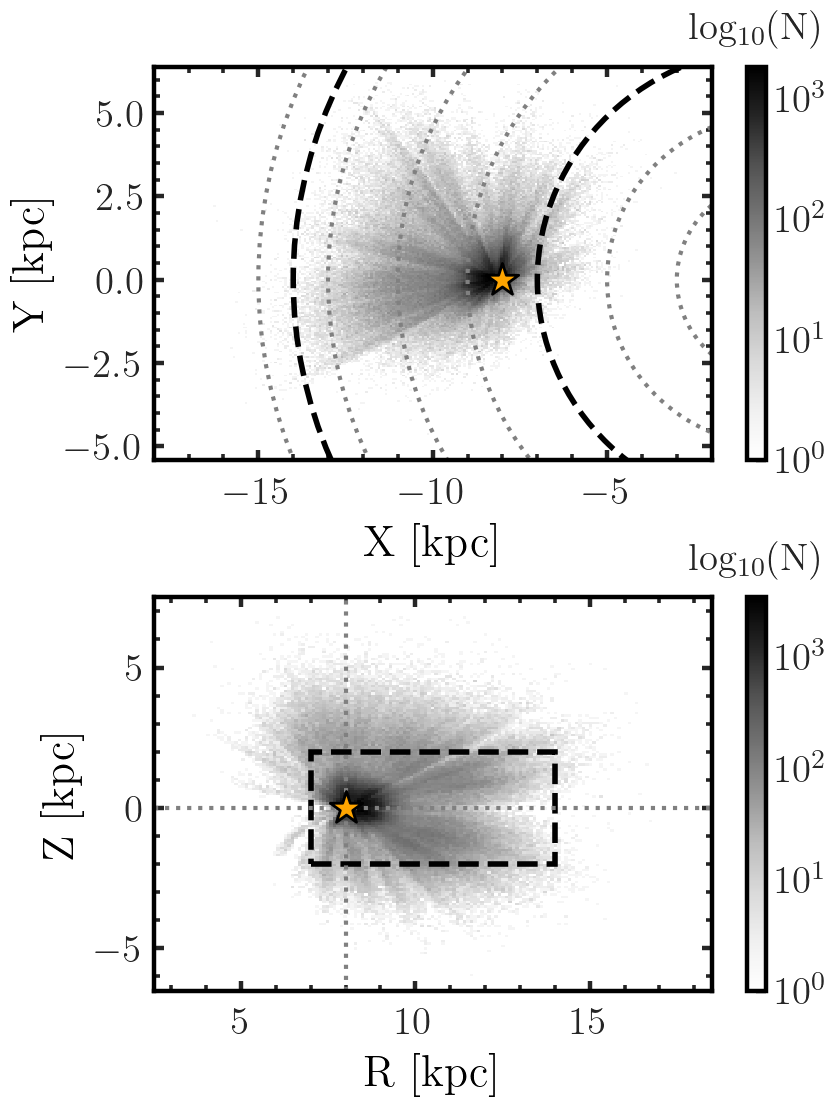}
    \caption{Spatial distribution for solar-like stars, shown as a face-on view  (X-Y plane; top panel), and an edge-on view (R-Z plane; bottom panel) of the Galaxy. The color-coding represents the logarithm of the number density. For reference, the solar position is denoted by an orange star. The black dashed lines represent the range within $7 \leq {\rm{R}} \leq 14$ kpc and $ |{\rm{Z}}| \leq 2 $ kpc.}
    \label{fig:xyz_density}
\end{figure}

\begin{figure}
	\centering
	\includegraphics[width=0.85\columnwidth]{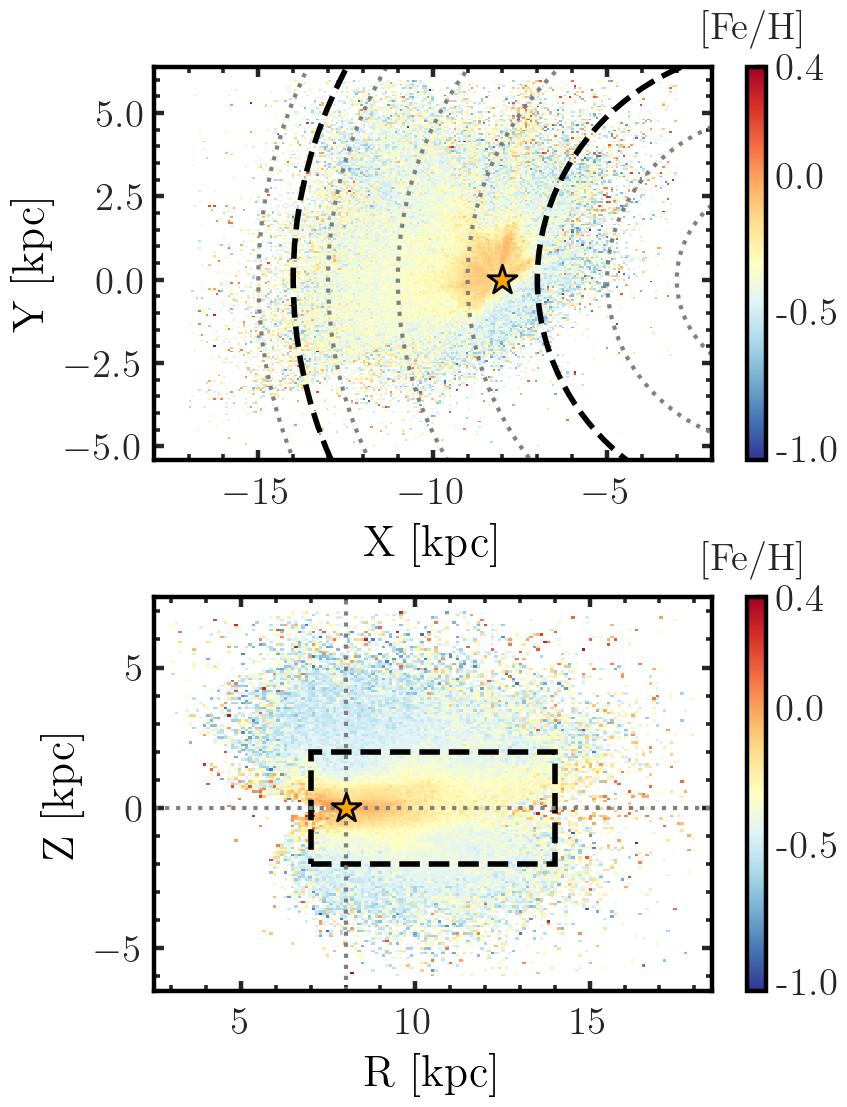}
    \caption{Spatial distribution of metallicity for solar-like stars. This figure is analogous to Fig. \ref{fig:xyz_density}, except the color-coding represents the median \feh \space in each bin.}
    \label{fig:xyz_metal}
\end{figure}

Based on our distances, we calculated the  galactocentric Cartesian coordinates using \texttt{astropy}.\footnote{\url{https://docs.astropy.org/en/stable/coordinates/index.html}} In Fig. \ref{fig:xyz_density} we show the X-Z (top panel) and R-Z (bottom panel) spatial distribution of the solar-like stars in our results. From these we selected a subsample of 458,061 stars within $7 \leq R \leq 14$ kpc and $ |Z| \leq 2 $ kpc to investigate the metallicity gradients. In Fig. \ref{fig:xyz_metal} we display the metallicity distribution in the same planes as Fig. \ref{fig:xyz_density}. The top panel shows that the inner Galaxy has a higher metallicity than the outer regions (indicating a negative radial metallicity gradient), while the bottom panel reveals that metallicity decreases as $|Z|$ increases (indicating a negative vertical metallicity gradient).

For this work  we chose a single linear model for both radial and vertical metallicity gradients. The radial metallicity gradient is defined as
\begin{equation}
	\label{eq:radial_grad}
	\feh_{R} = \frac{\Delta\feh}{\Delta R} R + b_R,
\end{equation}
where $\feh_{R}$ is the metallicity at a particular galactocentric radius $R$, $\Delta \feh / \Delta R$ is the radial metallicity gradient, and $b_R$ is the intercept. 
The vertical metallicity gradient is defined as
\begin{equation}
	\label{eq:vertical_grad}
	\feh_{Z} = \frac{\Delta \feh}{\Delta Z} Z + b_Z,
\end{equation}
where $\feh_Z$ is the metallicity at a particular absolute vertical height $|Z|$, $\Delta \feh / \Delta Z$ is the vertical metallicity gradient, and $b_Z$ is the intercept. 
However, since the metallicity gradients are not the main focus of this work, we applied a simple linear least-squares regression to estimate the gradients as well as their uncertainties ($\sigma_{\Delta \feh / \Delta R}$, $\sigma_{b_R}$; $\sigma_{\Delta \feh / \Delta Z}$, $\sigma_{b_Z}$) using the \texttt{scipy} library.\footnote{\url{https://docs.scipy.org/doc/scipy-1.14.0/reference/generated/scipy.stats.linregress.html}} To calculate the radial metallicity gradient, we divided our sample into equally spaced vertical bins with a width of 0.2 kpc, covering the range $0 < |Z| < 2$ kpc. Similarly, for the vertical metallicity gradient, the stars were divided into equally spaced radial bins from $7 < R < 14$ kpc, each with a bin width of 1 kpc.

\begin{figure}
	\centering
	\includegraphics[width=0.97\columnwidth]{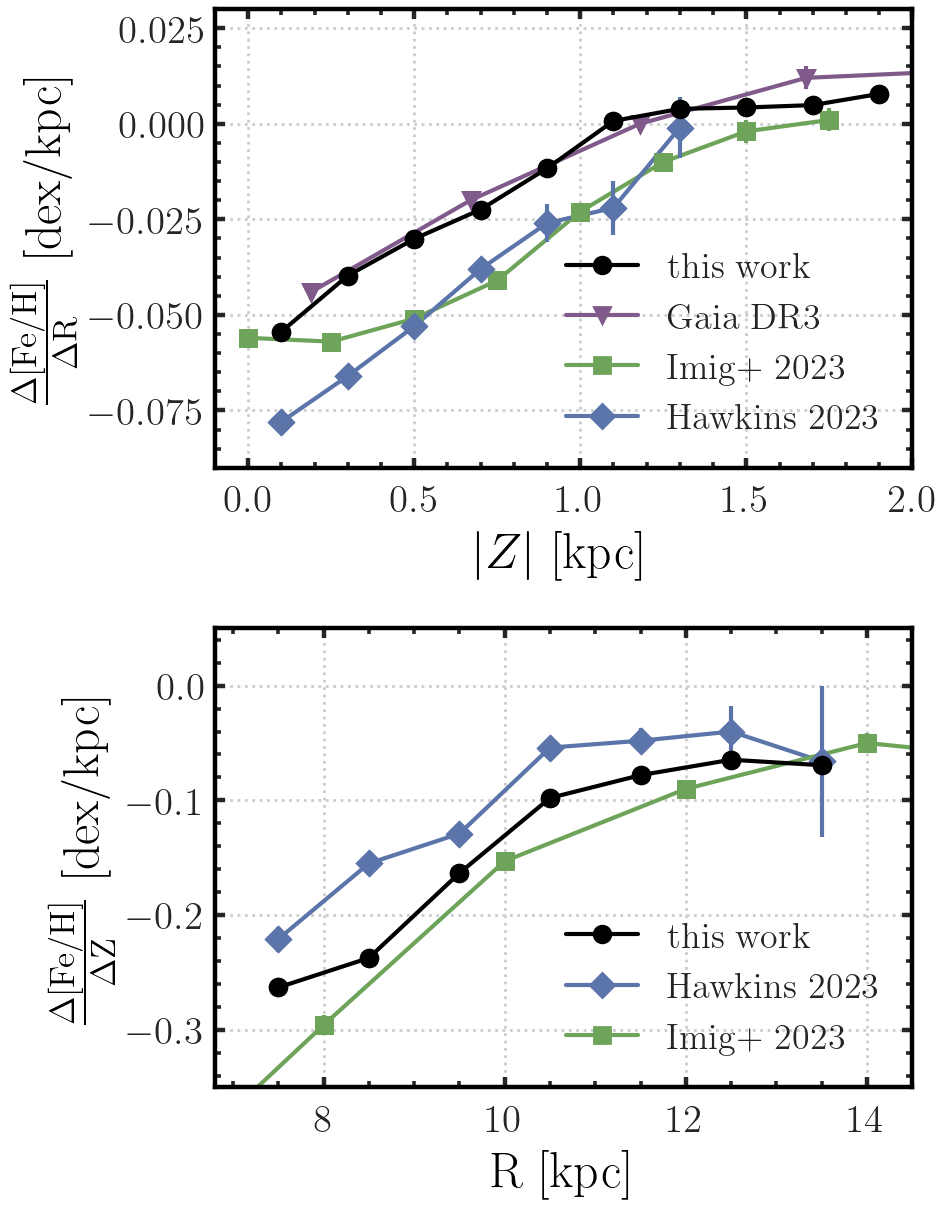}
    \caption{Radial and vertical metallicity gradients based on solar-like stars. Top panel: Radial metallicity gradient $\Delta \feh / \Delta R$ as a function of absolute height ($|Z|$) from the Galactic mid-plane for solar-like stars in this study (black circles, line). Bottom panel: Vertical metallicity gradient $\Delta \feh / \Delta Z$ as a function of galactocentric radius ($R$) (black circles, line). For comparison, we also show the radial metallicity gradient measured by \citet{gaia_collaboration_metal_2023} (purple triangles, line), \citet{imig_tale_2023} (green squares, line), and \citet{hawkins_chemical_2023} (blue diamonds, line). The data points are from Table \ref{tab:radial_grad} and Table \ref{tab:vertical_grad}.}
    \label{fig:comp_grad}
\end{figure}

The results are summarized in Tables \ref{tab:radial_grad} and \ref{tab:vertical_grad}. The radial metallicity gradient ranges from $-0.05$ to $0.0\ {\rm{dex\ kpc^{-1}}}$, while the vertical metallicity gradient ranges from $-0.26$ to $-0.07\ {\rm{dex\ kpc^{-1}}}$. These results are consistent with previous studies \citep{gaia_collaboration_metal_2023, hawkins_chemical_2023, imig_tale_2023}, as shown in Fig. \ref{fig:comp_grad}. In the top panel our result show the best agreement with those of \citet{gaia_collaboration_metal_2023}, which is based on FGK dwarfs and giants from the Gaia DR3 survey. \citet{imig_tale_2023} and \citet{hawkins_chemical_2023} used red giants from APOGEE and OBAF-type stars from LAMOST as tracers, respectively, and found steeper gradients compared to ours. These differences likely arise from the distinct tracer populations and observational surveys used. In the bottom panel, our results are also generally consistent with those of  \citet{hawkins_chemical_2023} and \citet{imig_tale_2023}. The discrepancies between these studies can be attributed to variations in binning methods and sample sizes. Additionally, it can be seen that the radial and vertical gradients generally flatten at larger distances from the Galactic center and greater vertical heights, respectively, which aligns well with previous findings. Our results for the metallicity gradients serve as a simple demonstration, illustrating that our distance and stellar parameter estimations have properties similar to what is expected based on our knowledge of metallicity distributions of Milky Way components.
\setlength{\tabcolsep}{3.2pt}
\begin{table}
    \centering
	\caption{Radial metallicity gradients for $|Z|$ bins.}
    \label{tab:radial_grad}
    \begin{tabular}{cccccc} 
		\hline \hline 
		\noalign{\smallskip}
         $|Z|$&  $\Delta \feh / \Delta R$& $\sigma_{\Delta \feh / \Delta R}$&  $b_R$&  $\sigma_{b_R}$& N\\ 
         (kpc)&  ($\rm{dex\ kpc^{-1}}$)&  ($\rm{dex\ kpc^{-1}}$)&  (dex)&  (dex)& \\ 
		 \noalign{\smallskip}
		 \hline  
		 \noalign{\smallskip}
         0.1&  $-0.054$&  0.001&  0.367&  0.008& 110 637\\ 
         0.3&  $-0.040$&  0.001&  0.192&  0.007& 111 861\\ 
         0.5&  $-0.030$&  0.001&  0.062&  0.007& 72 406\\ 
         0.7&  $-0.022$&  0.001&  $-0.051$&  0.008& 46 237\\ 
         0.9&  $-0.011$&  0.001&  $-0.202$&  0.008& 31 200\\ 
         1.1&  $0.000$&  0.001&  $-0.360$&  0.009& 23 342\\  
         1.3&  $0.003$&  0.001&  $-0.408$&  0.010& 19 706\\ 
         1.5&  $0.004$&  0.001&  $-0.434$&  0.011& 16 660\\ 
		 1.7& $0.005$& 0.001& $-0.456$& 0.011&14 009\\ 
 		 1.9& $0.008$& 0.001& $-0.497$& 0.011&12 003\\ 
		\noalign{\smallskip}
		\hline \hline
    \end{tabular}
\end{table}
\setlength{\tabcolsep}{3.2pt}
\begin{table}
    \centering
	\caption{Vertical metallicity gradients for $R$ bins.}
    \label{tab:vertical_grad}
    \begin{tabular}{cccccc} 
		\hline  \hline
		\noalign{\smallskip}
         $R$&  $\Delta \feh / \Delta Z$&  $\sigma_{\Delta \feh / \Delta Z}$&  $b_Z$&  $\sigma_{b_Z}$& N\\ 
         (kpc)&  ($\rm{dex\ kpc^{-1}}$)&  ($\rm{dex\ kpc^{-1}}$)&  (dex)&  (dex)& \\ 
		 \noalign{\smallskip}
		 \hline  
		 \noalign{\smallskip}
         7.5&  $-0.263$&  0.002&  $-0.044$&  0.001& 67 685\\   
         8.5&  $-0.237$&  0.001&  $-0.073$&  0.001& 247 688\\  
         9.5&  $-0.163$&  0.001&  $-0.124$&  0.001& 67 250\\   
         10.5&  $-0.098$&  0.002&  $-0.223$&  0.002& 38 715\\ 
         11.5&  $-0.078$&  0.002&  $-0.283$&  0.003& 24 062\\ 
         12.5&  $-0.065$&  0.003&  $-0.299$&  0.004& 9 809\\  
         13.5&  $-0.069$&  0.006&  $-0.293$&  0.008& 2 852\\ 
		 \noalign{\smallskip}
		 \hline \hline
    \end{tabular}
\end{table}

\section{Discussion and Conclusion}
\label{sec:discussion and conclusion}
\subsection{Feature importance}
\label{sec:feature importance}
To examine which parts of the spectra the CNN is weighting when predicting stellar parameters, we analyzed the feature importance through Shapley values. Using the Python package \texttt{SHAP} (Shapley Additive Explanations) \citep{shap}, we calculated the Shapley values of each feature for our CNN model, where each feature corresponds to a 1\AA \space wavelength. 

\begin{figure*}
	\centering
	\includegraphics[width=\textwidth]{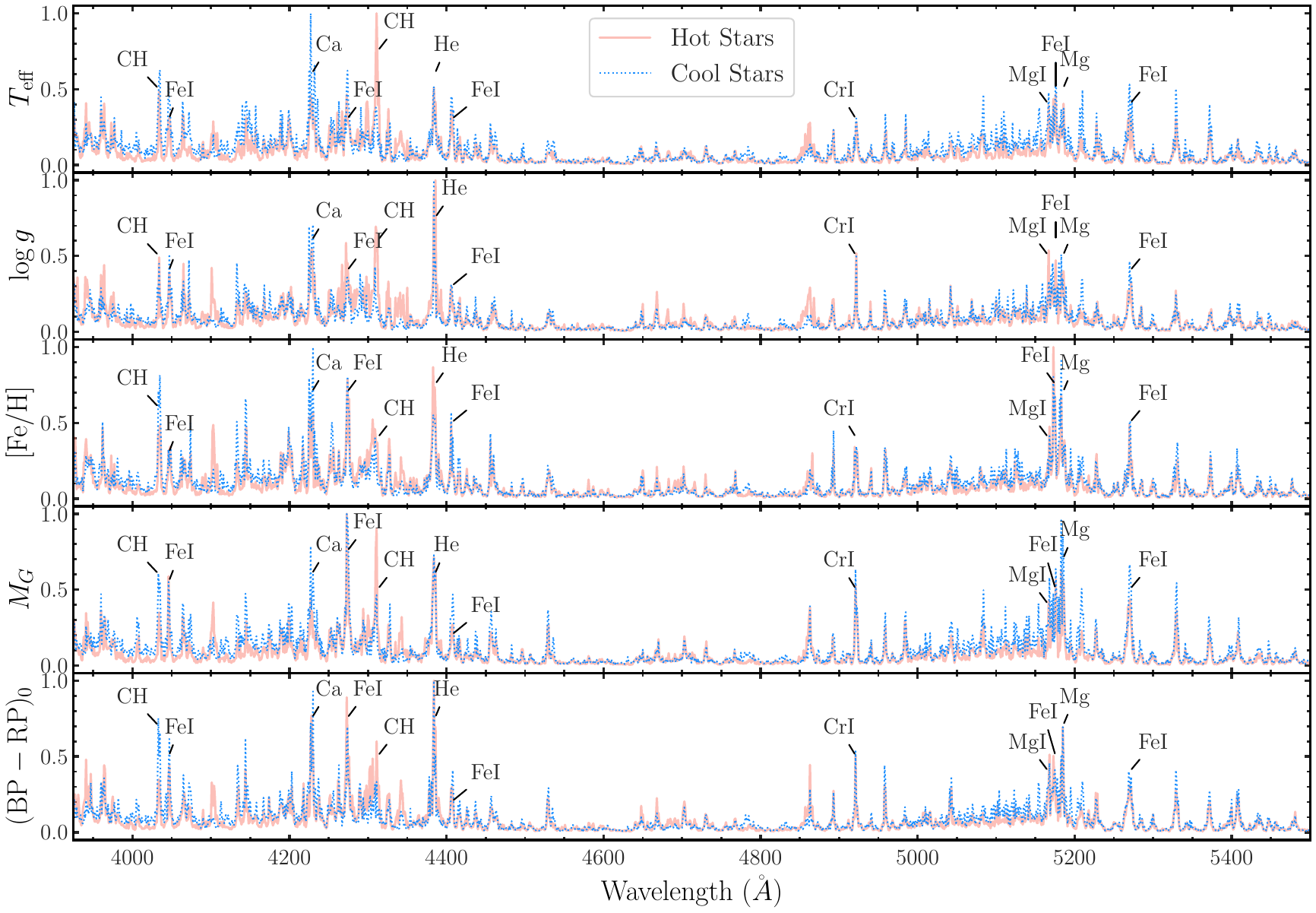}
    \caption{Normalized average absolute Shapley values of five output parameters obtained from the trained CNN model  shown over the input wavelength range of $3925 - 5500$\AA. We compare hot stars ($\teff > 5000$ K, shown as pink lines) and cool stars ($\teff \leq 5000$ K, shown as blue dotted lines). The wavelengths where strong features appear are annotated with their corresponding elements.}		
    \label{fig:shapteff}
\end{figure*}

\begin{figure*}
	\centering
	\includegraphics[width=\textwidth]{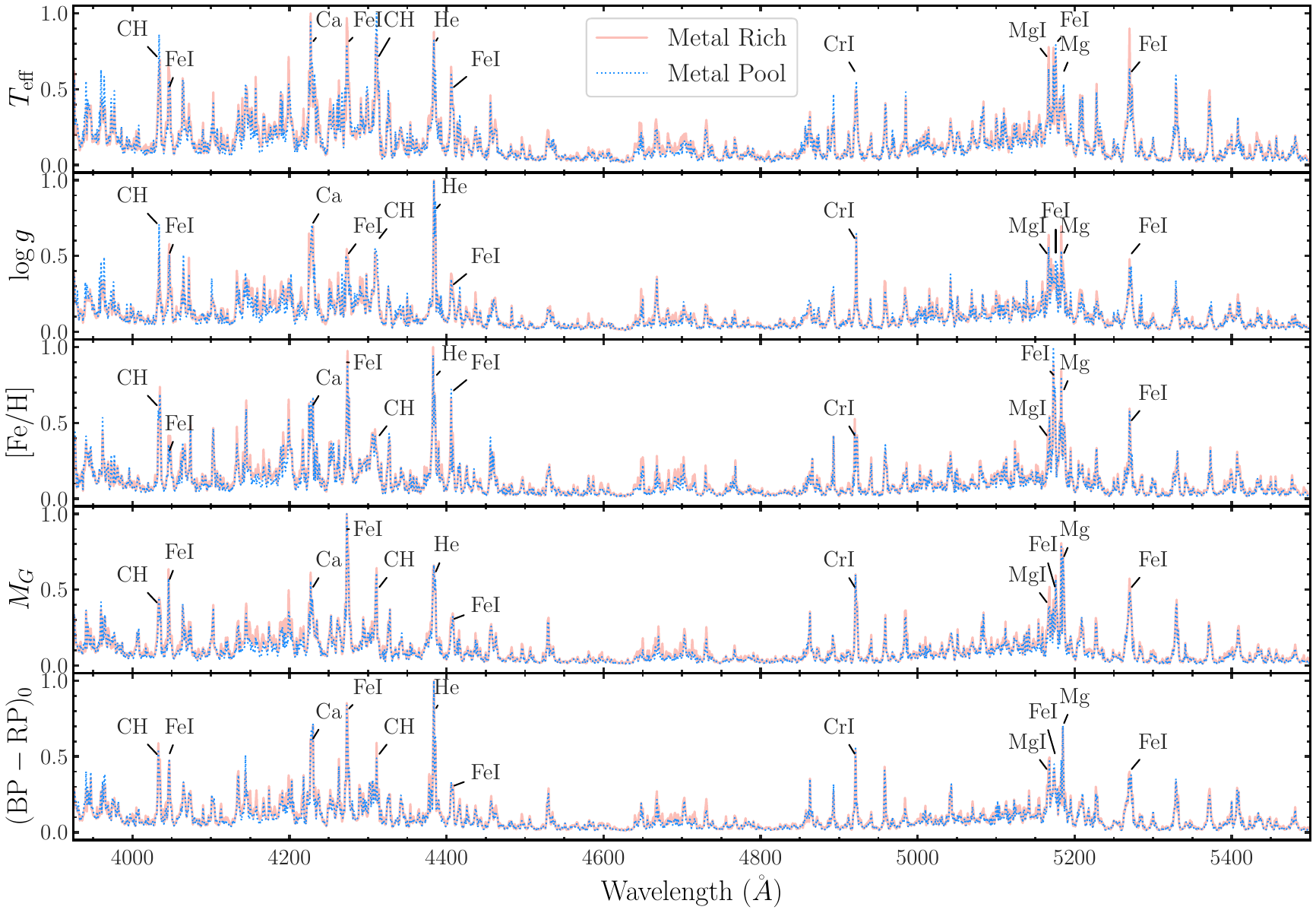}
    \caption{Similar to Fig. \ref{fig:shapteff}, but the comparison is made between metal-rich stars ($\feh \geq 0.0$ dex, shown as pink lines) and metal-poor stars ($\feh < 0.0$ dex, shown as blue dotted lines).}
    \label{fig:shapmetal}
\end{figure*}

We calculated the Shapley values for 2,000 stars. To quantify the impact, we computed the normalized average absolute Shapley values for each parameter, where larger values at a given wavelength indicate a greater influence on the parameter predictions. Comparisons are then made between hot ($\teff > 5000$ K) and cool stars ($\teff \leq 5000$ K), and between metal-poor stars ($\feh < 0.0$ dex) and metal-rich stars ($\feh \geq 0.0$ dex), as shown in Figs. \ref{fig:shapteff} and \ref{fig:shapmetal}. Only the results for the wavelength range of $3925-5500$\AA \space are demonstrated, as the $5500-8800$\AA \space range does not show significant importance.
A few notable features include the following:
\begin{enumerate}
    \item[(i)]{The helium line plays a crucial role in the determination of surface gravity and color, as well as the metallicity estimation for hot stars.}
    \vspace{0.5em}
    \item[(ii)]{Atomic metal lines, such as FeI, MgI, CrI, Ca, and Mg, significantly contribute to the determination of temperature and metallicity throughout our stellar parameter range.}
    \vspace{0.5em}
    \item[(iii)]{Spectral lines that substantially impact a particular parameter also influence other parameters, although their significance may vary. For instance, the FeI line at 4272\AA \space has a pronounced effect on absolute magnitude, but a weaker influence on surface gravity.}
    \vspace{0.5em}
    \item[(iv)]{Unidentified or weaker lines, including those in the 5500--8800\AA \space range that may not be prominent individually, collectively contribute significantly to the measurement of stellar parameters. Therefore, employing full spectrum analysis might be preferable to using spectral indices.}
    \vspace{0.5em}
    \item[(v)]{In Fig. \ref{fig:shapteff} for hot stars, the CH line plays an important role in determining temperature and absolute magnitude, while the helium line contributes substantially to the estimation of metallicity.}
\end{enumerate}

\subsection{Conclusion}
\label{sec:conclusion}
In this work  we proposed a convolutional neural network model to derive distances and stellar parameters (\teff, \logg, and \feh) directly from LAMOST low-resolution spectra, combined with photometric data from Gaia. 
The model is trained on the LAMOST DR9 spectra, using APOGEE DR17 stellar parameters as well as Gaia DR3 for solar-like stars. For spectra with $\rm{S/N_g} > 10$, the test shows a scatter of 86 K for \teff, 0.07 dex for \logg, 0.06 dex for \feh, 0.25 mag for \absm, and 0.03 mag for \bprp. The distance predictions have a scatter of 65 pc, indicating a median fractional error of less than 4\% and a standard deviation of 8\%.

Using the CNN model, we determined the distance and stellar parameters of 521,424 solar-like stars from LAMOST DR9, thus compensating for those that lack precise distance measurements. The reliability of the distance estimation was evaluated by comparing it with three other studies: StarHorse, DistMass, and astroNN. The comparisons yield mean fractional errors of $7\%$, $-4\%$, and $-4\%$, respectively, indicating good agreement with these studies. Additionally, the accuracy of the stellar parameters is validated by comparison with APOGEE DR17 and GALAH DR3, showing minor biases and different degrees of scatter but overall consistency with these datasets.

Moreover, using a subsample of solar-like stars from our dataset, we calculated the radial and vertical metallicity gradients of the Milky Way. The radial gradient ranges from $-0.05 < \Delta \feh / \Delta R < 0.0\ {\rm{dex\ kpc^{-1}}}$ for stars located at $0 < |Z| < 2$ kpc, while the vertical gradient ranges from $-0.26 < \Delta \feh / \Delta Z < -0.07\ \rm{dex\ kpc^{-1}}$ for stars situated at $7 < R < 14$ kpc. These results align well with previous studies, indicating that our distance and metallicity measurements are reliable for Galactic structure studies.

In the future, we plan to extend this model to include a broader range of stellar types, beyond solar-like stars. This includes improving the determination of distances and stellar parameters for a larger number of stars in spectroscopic surveys. 

\begin{acknowledgements}

This work is supported by the National Natural Science Foundation of China (12261141689, 12273075), the National Key R\&D Program of China No. 2019YFA0405502, and the science research grants from the China Manned Space Project with NO.CMS-CSST-2021-B05. 

We sincerely thank the anonymous referee for the helpful comments. Yue-Yue Shen thanks Bo Zhang for the insightful suggestions with the paper.

This work has made use of data from the Guoshoujing Telescope (the Large Sky Area Multi-Object Fiber Spectroscopic Telescope LAMOST), a National Major Scientific Project built by the Chinese Academy of Sciences. Funding for the project has been provided by the National Development and Reform Commission. LAMOST is operated and managed by the National Astronomical Observatories, Chinese Academy of Sciences.

This work has made use of data from the European Space Agency (ESA) mission
{\it Gaia} (\url{https://www.cosmos.esa.int/gaia}), processed by the {\it Gaia}
Data Processing and Analysis Consortium (DPAC,
\url{https://www.cosmos.esa.int/web/gaia/dpac/consortium}). Funding for the DPAC
has been provided by national institutions, in particular the institutions
participating in the {\it Gaia} Multilateral Agreement.

Funding for the Sloan Digital Sky Survey IV has been provided by the Alfred P. Sloan Foundation, the U.S. Department of Energy Office of Science, and the Participating Institutions. 
SDSS-IV acknowledges support and resources from the Center for High Performance Computing  at the University of Utah. The SDSS website is www.sdss4.org.

SDSS-IV is managed by the Astrophysical Research Consortium for the Participating Institutions of the SDSS Collaboration including the Brazilian Participation Group, the Carnegie Institution for Science, Carnegie Mellon University, Center for Astrophysics | Harvard \& Smithsonian, the Chilean Participation Group, the French Participation Group, Instituto de Astrof\'isica de Canarias, The Johns Hopkins University, Kavli Institute for the Physics and Mathematics of the Universe (IPMU) / University of Tokyo, the Korean Participation Group, Lawrence Berkeley National Laboratory, Leibniz Institut f\"ur Astrophysik Potsdam (AIP),  Max-Planck-Institut f\"ur Astronomie (MPIA Heidelberg), Max-Planck-Institut f\"ur Astrophysik (MPA Garching), Max-Planck-Institut f\"ur Extraterrestrische Physik (MPE), National Astronomical Observatories of China, New Mexico State University, New York University, University of Notre Dame, Observat\'ario Nacional / MCTI, The Ohio State University, Pennsylvania State University, Shanghai Astronomical Observatory, United Kingdom Participation Group, Universidad Nacional Aut\'onoma de M\'exico, University of Arizona, University of Colorado Boulder, University of Oxford, University of Portsmouth, University of Utah, University of Virginia, University of Washington, University of Wisconsin, Vanderbilt University, and Yale University.

This work made use of the Third Data Release of the GALAH Survey (Buder et al. 2021). The GALAH Survey is based on data acquired through the Australian Astronomical Observatory, under programs: A/2013B/13 (The GALAH pilot survey); A/2014A/25, A/2015A/19, A2017A/18 (The GALAH survey phase 1); A2018A/18 (Open clusters with HERMES); A2019A/1 (Hierarchical star formation in Ori OB1); A2019A/15 (The GALAH survey phase 2); A/2015B/19, A/2016A/22, A/2016B/10, A/2017B/16, A/2018B/15 (The HERMES-TESS program); and A/2015A/3, A/2015B/1, A/2015B/19, A/2016A/22, A/2016B/12, A/2017A/14 (The HERMES K2-follow-up program). We acknowledge the traditional owners of the land on which the AAT stands, the Gamilaraay people, and pay our respects to elders past and present. This paper includes data that has been provided by AAO Data Central (datacentral.org.au).

The preparation of this work has made use of TOPCAT\citep{topcat}, NASA's Astrophysics Data System Bibliographic Services, as well as the open-source Python packages astropy\citep{astropy:2013, astropy:2018, astropy:2022}, NumPy\citep{numpy}, tensorflow\citep{tensorflow2015-whitepaper}, scipy\citep{2020SciPy-NMeth}, scikit-learn\citep{scikit-learn} and pandas\citep{mckinney-proc-scipy-2010}. The figures in this paper were produced with matplotlib\citep{matplotlib}.

\end{acknowledgements}





\bibliographystyle{aa} 
\bibliography{article.bib} 

\begin{appendix}
\onecolumn
\section{Dataset for CNN model}
\label{appendix-trainingset}
\begin{figure}[!htbp]
	\centering
	\includegraphics[width=\columnwidth]{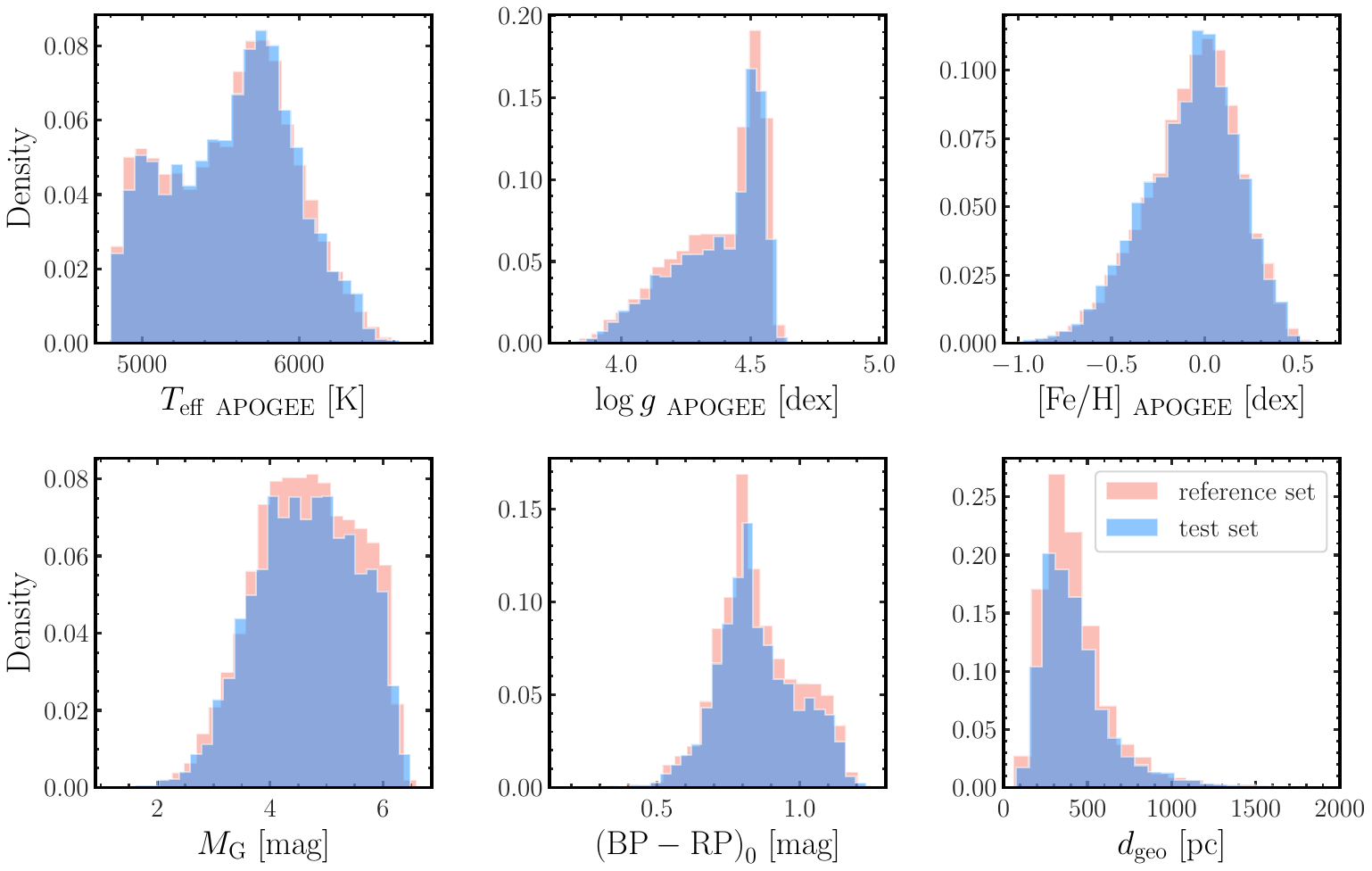}
    \caption{Parameter distribution of the dataset, with the reference and test sets colored pink and blue, respectively. The CNN model may underperform on stars in sparsely populated regions of the training set parameter space.}
    \label{fig:param-pdf}
\end{figure}

The parameter distribution of the dataset with a total of 18,573 stars is shown in Fig. \ref{fig:param-pdf}. The reference set consists of 14,858 stars, while the test set contains 3,715 stars.

\section{Pre-processing of the spectra}
\label{appendix-preprocessing}
\begin{figure}[!htbp]
	\centering
	\includegraphics*[width=\textwidth]{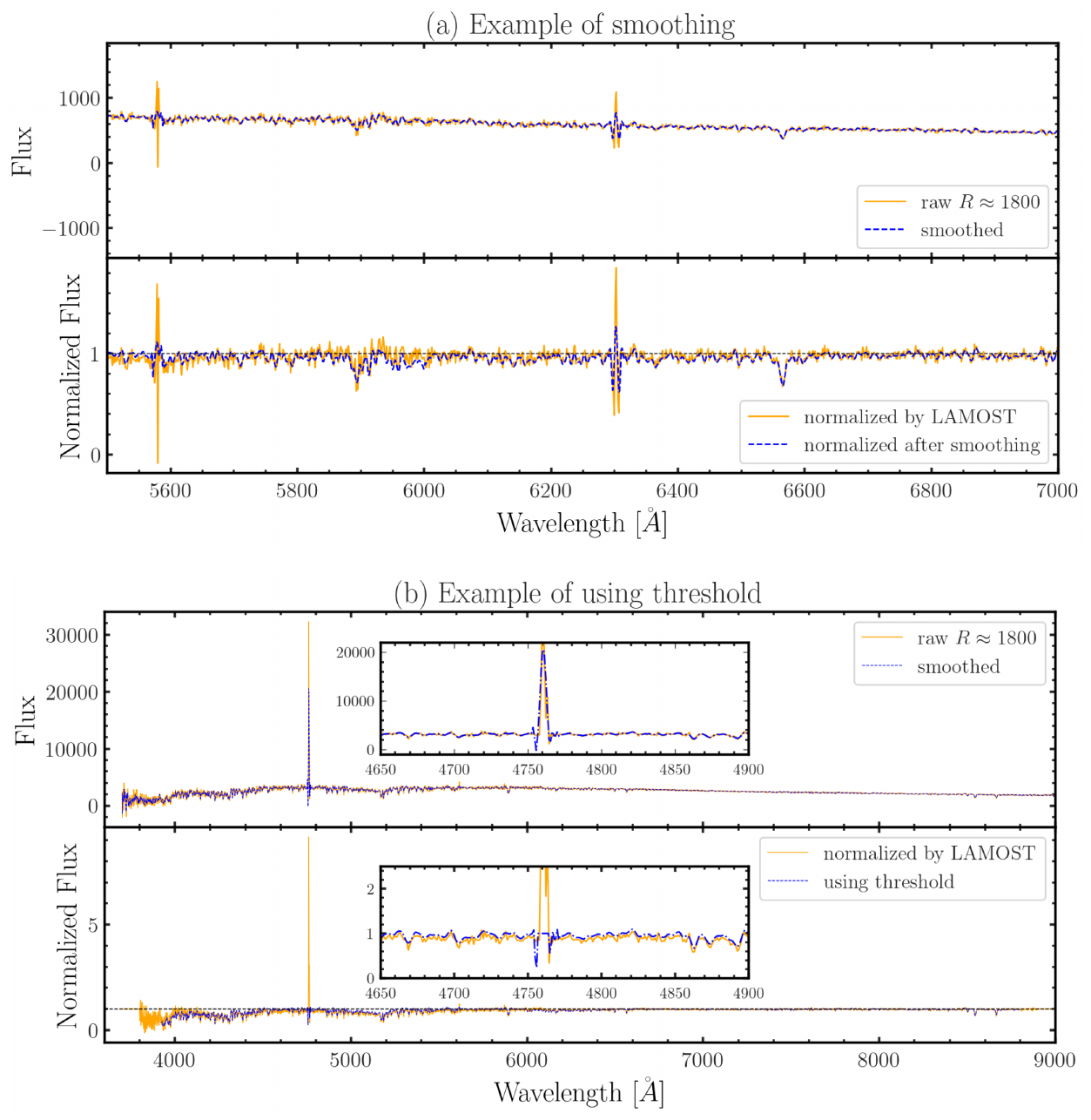}
	\caption{
		(a) Example of spectrum smoothing. The upper panel presents the raw spectrum (orange line) and the smoothed spectrum (blue dashed line). The lower panel shows the normalized spectrum from the LAMOST FITS file (orange line) and the normalized spectrum derived from the smoothed spectrum (blue dashed line).
		(b) Example of threshold processing for a spectrum with highly abnormal flux. The upper panel displays a raw spectrum with very high anomalous pixels (orange line) and its smoothed counterpart (blue dashed line). The lower panel shows the normalized spectrum from the LAMOST FITS file (orange line) and the spectrum corrected for anomalous pixels using a threshold (blue dashed line).}
	\label{fig:preprocess}
\end{figure}

Two examples of pre-processing are shown in Fig. \ref{fig:preprocess}.

\section{Model comparison}
\label{sec:model comparison}
The primary goal of this study is to solve a multi-label regression problem, which can be addressed by many Machine Learning (ML) algorithms. To compare our CNN model with other ML algorithms, we use the AutoGluon toolkit \citep{autogluon}, an open-source\footnote{\url{https://github.com/autogluon/autogluon}} AutoML framework. With AutoGluon, we can automatically train and tune several traditional models specifically on our dataset, including LightGBM boosted trees \citep{ke2017lightgbm}, CatBoost boosted trees \citep{dorogush2018catboost}, Random Forests, Extremely Randomized Trees, $k$-Nearest Neighbors, and multilayer perceptrons (MLP). This approach ensures a fair and comprehensive comparison by effectively optimizing each traditional model's hyperparameters.

To further elaborate, AutoGluon tackles a multi-label problem by breaking it down into multiple single-label regression tasks. The process is sequential: the model first predicts one label, then uses this prediction as an input for predicting the next label. Correlations between labels are accounted for by imposing an order on the labels, ensuring that the predictions for later labels incorporate the information from earlier ones. This sequential process captures the dependencies between labels more effectively.

We evaluated these models using the same test set and performance metric as our CNN model to enable model comparison, as presented in Table \ref{tab:modelcomp}. The CNN represents our approach, while the other models are automatically tuned by AutoGluon without any human intervention. LightGBMXT, LightGBM, and LightGBMLarge are three different versions of the LightGBM boosted model, each with different hyperparameters. NeuralNetFastAI and NeuralNetTorch are both MLPs with different implementations. KNeighborsUnif and KNeighborsDist are both $k$-Nearest Neighbors models, differing in their weight functions. For \teff \space prediction, the CNN outperforms all other models, achieving the smallest MSE value. The CNN also performs well for \logg, \feh, \absm, and \bprp. 
Although the CNN model does not exhibit the best overall performance, it remains competitive given its computational efficiency. It is evident that AutoGluon requires additional computation time and storage space to train and tune these various models. Furthermore, the precision differences to three decimal places are unlikely to significantly improve the results. In such cases, the CNN model is a better choice.
\begin{table*}[!htbp]
	\centering
	\caption{Comparison of our CNN model with other models.}
	\label{tab:modelcomp}
	\begin{tabular}{ccccccc}
		\hline \hline
		\noalign{\smallskip}
		Model           & \teff & \logg & \feh & \absm & \bprp & Weight mean MSE\\ 
		\noalign{\smallskip}
		\hline
		\noalign{\smallskip}
		CNN             & 7390      & 0.00497   & 0.00318  & 0.0649       & 0.00117        & 0.0685    \\ 
		LightGBMXT      & 7653      & 0.00409   & 0.00236  & 0.0531       & 0.0011         & 0.0606    \\ 
		LightGBM        & 7803      & 0.00409   & 0.00259  & 0.0558       & 0.00108        & 0.0619    \\ 
		LightGBMLarge   & 7815      & 0.0045    & 0.0033   & 0.0539       & 0.00109        & 0.0648    \\ 
		CatBoost        & 7925      & 0.00401   & 0.0023   & 0.0536       & 0.00117        & 0.0618    \\ 
		NeuralNetFastAI & 8671      & 0.00438   & 0.00165  & 0.0615       & 0.00136        & 0.0668     \\ 
		NeuralNetTorch  & 8313      & 0.00561   & 0.00323  & 0.0672       & 0.00158        & 0.0782    \\
		ExtraTreesMSE   & 8470      & 0.00628   & 0.0051   & 0.0607       & 0.00128        & 0.0803    \\ 
		RandomForestMSE & 8523      & 0.00586   & 0.00568  & 0.0614       & 0.00123        & 0.0804    \\ 
		KNeighborsUnif  & 12530     & 0.01385   & 0.02056  & 0.1644       & 0.00212        & 0.1891    \\ 
		KNeighborsDist  & 12607     & 0.01377   & 0.02015  & 0.1635       & 0.00211        & 0.1875    \\ 
		\noalign{\smallskip}
		\hline \hline
	\end{tabular}
	\tablefoot{Other models are automatically tuned by AutoGluon using the same training set. The performance values are evaluated using the test set. The first five columns show the MSE values of the five parameters under different models, demonstrating the parameter prediction ability for each parameter. The last column shows the weighted mean MSE of all five parameters, reflecting the overall performance of each model.}
\end{table*}

\section{Catalog of 521,424 solar-like stars}
\label{appendix-catalog}
Table \ref{tab:finalcatalog} presents the final catalog of this work, including the stellar parameters and distance of solar-like stars of LAMOST and Gaia. The complete catalog is publicly available through the China-VO Paper Data Repository and can be accessed at \url{https://nadc.china-vo.org/res/r101400/}. In addition, both the dataset and the trained model are accessible via Zenodo at doi: \href{https://zenodo.org/doi/10.5281/zenodo.13748129}{[10.5281/zenodo.13748129]}. The source code is hosted on GitHub at \url{https://github.com/sarashenyy/SolarDis}. 
\begin{table*}[!h]
	\centering
	\caption{The description of columns of the final catalog.}
	\label{tab:finalcatalog}
	\begin{tabular}{lccl} 
		\hline \hline
        \noalign{\smallskip}  
		Column&  Unit&Type& Description\\ 
		\noalign{\smallskip}
		\hline
		\noalign{\smallskip}
		obsid&  &integer& LAMOST observation identifier\\ 
		ra&  degree&double& right ascension of object\\ 
		dec&  degree&double& declination of object\\ 
		snrg&  &float& S/N at g band\\ 
		snrr&  &float& S/N at r band\\ 
		flag&  &short& flag=1 means existing bad pixels\\ 
		cnn\_teff&  K&double& effective temperature determined by this work\\ 
		cnn\_logg&  dex&double& surface gravity determined by this work\\ 
		cnn\_feh&  dex&double& Metallicity determined by this work\\ 
		cnn\_mg&  mag&double& absolute magnitude of Gaia G band determined by this work\\ 
		cnn\_bprp&  mag&double&BP - RP color of Gaia determined by this work\\ 
		cnn\_distance&  pc&double&distance determined by this work\\ 
		lasp\_teff&  K&float&effective temperature obtained by LASP\\ 
		lasp\_teff\_err&  K&float& effective temperature uncertainty obtained by LASP\\ 
		lasp\_logg& dex& float&surface gravity obtained by LASP\\ 
		lasp\_logg\_err&  dex&float&surface gravity uncertainty obtained by LASP\\ 
		lasp\_feh&  dex&float&metallicity obtained by LASP\\ 
		lasp\_feh\_err&  dex&float&metallicity uncertainty obtained by LASP\\ 
		lasp\_rv&  km/s&float&radial velocity obtained by LASP\\ 
		lasp\_rv\_err&  km/s&float&radial velocity uncertainty obtained by LASP\\ 
		gaia\_source\_id&  &long&source identifier in Gaia DR3\\ 
		parallax&  mas&double&parallax provided by Gaia DR3\\ 
		parallax\_over\_error&  &double&parallax divided by its standard error\\ 
		parallax\_zeropoint&  mas&double&parallax zero-point according to \citet{Lindegren2021}\\ 
		parallax\_correction&  mas&double&parallax after zero-point correction\\ 
		gaia\_g\_mean\_mag&  mag&double&G mag provided by Gaia DR3\\ 
		gaia\_bp\_rp&  mag&double&BP - RP color provided by Gaia DR3\\ 
		r\_med\_geo&  pc&double&distance from \citet{bailer-jones_estimating_2021}\\ 
		APOGEE\_ID&  &string&identifier in APOGEE\\ 
        \noalign{\smallskip}  
		\hline \hline
	\end{tabular}
\end{table*}

\end{appendix}

\end{document}